\documentclass[%
 reprint,
superscriptaddress,
 amsmath,amssymb,
 aps,
prb,
]{revtex4-1}

\usepackage{caption}
\usepackage{graphicx}
\usepackage{dcolumn}
\usepackage{bm}
\usepackage{xcolor}

\begin{document}

\preprint{APS/123-QED}

\title{Revisiting Jahn--Teller Transitions in Correlated Oxides with Monte Carlo Modeling}

\author{Liam A. V. Nagle-Cocco}
 \email{lnc@slac.stanford.edu}
\affiliation
{Stanford Synchrotron Radiation Lightsource, SLAC National Accelerator Laboratory, Menlo Park, California, 94025, United States of America}


\author{Andrew L. Goodwin}
\affiliation{Inorganic Chemistry Laboratory, Department of Chemistry, University of Oxford, Oxford, OX1 3QR, United Kingdom.}

	\author{Clare P. Grey}
	\affiliation{Yusuf Hamied Department of Chemistry, University of Cambridge, Cambridge, CB2 1EW, United Kingdom.}
	
	\author{Si\^an E. Dutton}
	\affiliation{Cavendish Laboratory, University of Cambridge, JJ Thomson Avenue, Cambridge, CB3 0US, United Kingdom.}

\date{\today}

\begin{abstract}
Jahn--Teller (JT) distortions are a key driver of physical properties in many correlated oxide materials. Cooperative JT distortions, in which long-range orbital order reduces the symmetry of the average structure, are common in JT-distorted materials at low temperatures. This long-range order will often melt on heating, \textit{via} a transition to a high-temperature state without long-range orbital order. The nature of this transition has been observed to vary with different materials depending on crystal structure; in LaMnO$_3$ the transition has generally been interpreted as order-disorder, whereas in layered nickelates $A$NiO$_2$ ($A$=Li,Na) there is a displacive transition. 
However, recent theoretical work has suggested that previous evidence for order-disorder may in fact be a consequence of phonon anharmonicity, rather than persistence of JT distortions. 
In this work, we run Monte Carlo simulations with a simple Hamiltonian that is modified to include terms dependent on the JT amplitude $\rho$, which is allowed to vary within the simulation \textit{via} the Metropolis algorithm. 
Our simulations yield distributions of JT amplitudes consistent with displacive rather than order-disorder behaviour for both perovskites and layered nickelates. 
We also find significant differences between the transition observed for perovskites compared with layered nickelates, which we attribute to differing extensivity of configurational entropy on the two lattices, \textcolor{black}{suggesting a crucial role for} lattice geometry in determining behaviour. 
\end{abstract}


\maketitle

\section{Introduction}

Transition metal oxide octahedra with electronic configurational degeneracy are susceptible to a Jahn--Teller (JT) distortion~\cite{Jahn1937StabilityDegeneracy,Opik1957StudiesProblem,Longuet-Higgins1958StudiesProblem,Kanamori1960CrystalCompounds,Gehring1975Co-operativeEffects,Halcrow2013JahnTellerMaterials,Goodenough1998Jahn-TellerSolids}, causing their shape to deviate from that of a regular octahedron. 
Such distortions most commonly manifest as a linear superposition of the $E_g(Q_2,Q_3)$ van Vleck modes~\cite{VanVleck1939TheXY6,Nagle-Cocco2024VanVanVleckCalculator}, where $Q_2$ and $Q_3$ are planar rhombic and tetragonal distortions, respectively, although the tetragonal $Q_3$ component tends to dominate. 
JT distortions have been observed in both molecules~\cite{Halcrow2013JahnTellerMaterials} and crystals~\cite{Goodenough1998Jahn-TellerSolids}. 
The JT distortion is relevant to many phenomena including superconductivity in the cuprates~\cite{Fil1992Lattice-mediatedCuprates,Keller2008Jahn-TellerSuperconductivity,Bussmann-Holder2022SuperconductivityPolaron}, spin-orbit ordering~\cite{Khomskii2021OrbitalOpportunities}, and ionic mobility~\cite{Kim2015AnomalousBatteries,Li2016Jahn-TellerBatteries}, and can lead to structural transitions on cycling in battery electrode materials~\cite{Kim2015AnomalousBatteries,Li2016Jahn-TellerBatteries,Choi2019K0.54Co0.5Mn0.5O2:Batteries}. 

\begin{figure*}[t]
    \includegraphics[scale=1]{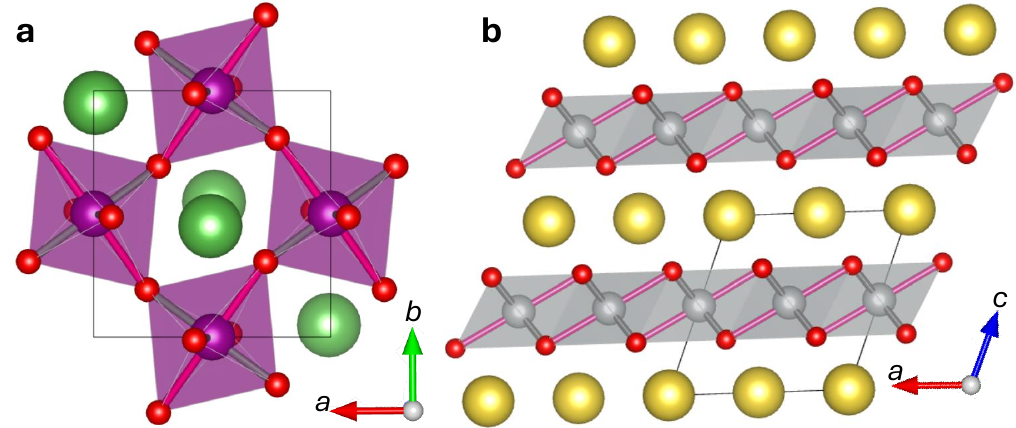}
    \captionsetup{list=no} 
    \caption{
        \label{JT-ordering-from-literature}
        The room-temperature crystal structures and JT-ordering of (a) many JT-ordered perovskites such as KCuF$_3$ and LaMnO$_3$~\cite{Khomskii2021OrbitalOpportunities}, corresponding to phase 5 of the phase diagram of Ahmed and Gehring~\cite{Ahmed2005TheModel}, and (b) layered nickelates such as NaNiO$_2$ and stoichiometric LiNiO$_2$~\cite{Dyer1954AlkaliMNiO2,Phillips2025Collinearsub2/sub}. 
        The stacking in (a) is commonly termed C-type orbital order. 
        Pink metal-oxygen bond lengths indicate JT-elongated bonds. 
        Only octahedra associated with JT-active sites are displayed. 
        In the perovskites, purple sites are the JT-distorted site (i.e. Mn, Cu, Cr) and green sites are the \textit{A}-site (i.e. La, K). 
        In the nickelates, grey sites are Ni$^{3+}$ and yellow sites are the alkali metal. 
    }
\end{figure*}

Within solids, the axes of octahedral elongation can correlate \textit{via} orbital ordering over macroscopic length-scales as a cooperative JT (cJT) distortion, typically manifesting as a decrease in unit cell symmetry~\cite{Gehring1975Co-operativeEffects,Goodenough1998Jahn-TellerSolids}. 
For instance, the perovskites K\textit{M}F$_3$ (\textit{M}=Cr,Cu)~\cite{Zhou2011Jahn-TellerPressure,Margadonna2006CooperativePerovskite}, LaMnO$_3$~\cite{Rodriguez-Carvajal1997TheLaMnO3}, and [(CH$_3$)$_2$NH$_2$]Cu(HCOO)$_3$~\cite{Scatena2021Pressure-inducedFreedom} exhibit a cJT distortion with alternating \textit{M}O$_6$ (\textit{M}=Cu,Cr,V,Mn) octahedra elongated along two orthogonal directions within the $xy$-plane, as shown in Figure~\ref{JT-ordering-from-literature}(a). 
Such planes generally show C-type stacking along the $c$-direction, in which each plane has identical ordering~\cite{Rodriguez-Carvajal1997TheLaMnO3,Ren1998Temperature-inducedCrystal,Blake2009CompetitionHoVO3,Zhou2011Jahn-TellerPressure}. 
Interestingly, similar orbital ordering driven by $t_{2g}$ orbitals is seen in the vanadate perovskites \textit{A}VO$_3$ (\textit{A}=Ho,La,Y)~\cite{Blake2009CompetitionHoVO3,Ren1998Temperature-inducedCrystal,Blake2001TransitionYVO3,Blake2002NeutronYVO3}, although on cooling to low temperatures, some (such as LaVO$_3$ and YVO$_3$) have been shown to exhibit out-of-phase stacking along the $c$-direction~\cite{Bordet1993StructuralK,Blake2001TransitionYVO3,Blake2002NeutronYVO3} [SI Figure~\ref{SI_G-type_from-lit.pdf}]. 
The flexibility of the perovskite structure means that other kinds of ordering can also be exhibited; for instance, Ba$_{0.5}$La$_{0.5}$CoO$_3$ exhibits a collinear elongational cJT ordering of the CoO$_6$ octahedra when the La/Ba is disordered~\cite{Fauth2002IntermediateDistortions} but a compressive collinear ordering when the La/Ba is ordered~\cite{Nakajima2005NewEffect} [Figure~\ref{Ba0.5La0.5CoO3_order_SI}]. 

cJT distortions also occur in the layered triangular-lattice nickelate NaNiO$_2$~\cite{Dyer1954AlkaliMNiO2,Dick1997TheScattering,Chappel2000StudyESR,Sofin2005NewNaNiO2,Nagle-Cocco2022PressureNaNiO2,Nagle-Cocco2024DisplaciveNaNiO2,Nagle-Cocco2025Dome-likeNaNiO2}, where diffraction shows a collinear ordering in which NiO$_6$ octahedra within each layer elongate along a common axis that is preserved from layer to layer [Figure~\ref{JT-ordering-from-literature}(b)]. 
By contrast, the electronic structure of LiNiO$_2$ has long remained under debate, with the as-synthesised material typically not showing a comparable cJT distortion~\cite{Dyer1954AlkaliMNiO2,bianchi2001synthesis,Chung2005LocalDiffraction,petit2006ground}. 
This has been attributed to suppression of long-range orbital order by Ni/Li site mixing and related defects~\cite{petit2006ground,lin2022defect,Genreith-Schriever2024Jahn-TellerDiffraction,Phillips2025Collinearsub2/sub}. 
Experimental evidence from neutron total scattering~\cite{Chung2005LocalDiffraction} and X-ray absorption spectroscopy~\cite{Rougier1995Non-cooperativeStudy} shows evidence that JT distortion likely occurs locally in conventional LiNiO$_2$, without long-range order. 
An alternative explanation is charge disproportionation, for which there is evidence from both computational studies~\cite{Chen2011ChargeStudy,Foyevtsova2019LiNiO2Glass,Poletayev2025Temperature-dependentLiNiO2} and X-ray absorption spectroscopy measurements~\cite{Green2020EvidenceSpectroscopy,Takegami2024ValenceDisproportionation,Poletayev2025Temperature-dependentLiNiO2} by analogy to layered silver nickelate~\cite{wawrzynska2007orbital,kang2007valence} and the nickelate perovskites~\cite{garcia1994neutron,mizokawa2000spin,garcia2009structure}. 
More recent studies of carefully prepared, stoichiometric LiNiO$_2$ have reported evidence for a collinear cJT distortion similar to that of NaNiO$_2$~\cite{Phillips2025Collinearsub2/sub}. 

Beyond perovskites and layered nickelates, JT-driven orbital order occurs in a range of other materials. 
The K$_2$PtCl$_6$-type material (NO)$_2$VCl$_6$, which has a compressive, not elongational, distortion, exhibits a collinear cJT distortion within the $ab$-plane, but the axis of octahedral compression alternates with layer stacking~\cite{Henke2003CrystalNO2VCl6}.
The spinels Mn$_3$O$_4$~\cite{Baron1998TheMn2+FeMn23+04}, LiMn$_2$O$_4$~\cite{Yamaguchi1998Jahn-TellerSpectroscopy}, MgMn$_2$O$_4$~\cite{Yokozaki2021EffectBattery}, and ZnMn$_2$O$_4$~\cite{Patra2019StructuralZnMn2O4} [Figure~\ref{SI_spinel_ZnMn2O4_literature.pdf}], wolframite CuWO$_4$~\cite{Schofield1997DistortionCu1-xZnxWO4}, Prussian Blue analogs such as CuPt(CN)$_6$~\cite{Harbourne2024LocalAnalogues} and K$_2$Cu[Fe(CN)$_6$]~\cite{Cattermull2022UncoveringK2CuFeCN6}, and Ruddlesden–Poppers like \textit{A}$_2$CuF$_4$ (\textit{A}=Na,K)~\cite{Herdtweck1981RontgenographischeK3Cu2F7,Hiley2025Pressure-inducedNa2CuF4}, all exhibit simple collinear cJT distortions, although a pressure-induced change in cJT ordering has been observed the spinel in Na$_2$CuF$_4$~\cite{Hiley2025Pressure-inducedNa2CuF4}. 
The distorted trirutile CuSb$_2$O$_6$~\cite{Nakua1991CrystalCuSb2O6} also exhibits intra-planar collinear ordering, but alternating $ab$-planes exhibit different axes of elongation. 
There are clearly a wide range of possible orbital orderings which may be energetically favourable, depending on the geometry of the octahedral interactions and the JT-active cation. 

Heating materials with a cJT distortion will often result in a symmetry-raising transition. 
In many perovskites, an order-disorder transition has been reported, in which JT distortions persist locally in octahedra even when cJT distortions are suppressed. 
For example, the canonical orbital-ordered material LaMnO$_3$ shows the melting of long-range orbital order around $T_\mathrm{JT}\approx750$\,K~\cite{Granado2000Order-disorderStudy,Chatterji2003VolumeTransition,zhou2003orbital,Qiu2005OrbitalLaMnO3,Kharlamova2009PhaseLaMnO3,Chatterji2004OrbitalRegion,Souza2004LocalTransition,Mondal2011Current-drivenLaMnO3,Ramirez2011StructuralLaMnO3,Thygesen2017LocalLaMnO3,Tran2019PhaseMethod,Saint-Paul2004Soft-acoustic3,tragheim2025interplay}. 
Local structure probes such as neutron total scattering~\cite{Qiu2005OrbitalLaMnO3,Thygesen2017LocalLaMnO3} and extended X-ray absorption fine structure (EXAFS)~\cite{Souza2004LocalTransition} have typically been interpreted as showing that distortions persist locally within octahedra, and correlate over short lengthscales. 
KCrF$_3$ shows a similar transition in crystal symmetry below $\sim$973\,K~\cite{Margadonna2007HighPerovskite}, which is likewise proposed to be order-disorder~\cite{Margadonna2007HighPerovskite} by analogy to LaMnO$_3$. 
Surprisingly, the isostructural KCuF$_3$ shows no sign of the melting of orbital order below its decomposition temperature~\cite{Marshall2013UnusualDiffraction}. 

Besides perovskites, there are reports of order-disorder transitions in other systems too. 
The spinel LiMn$_2$O$_4$, although a complex case due to mixed-valence Mn, exhibits a subtle tetragonal$\rightarrow$cubic JT transition which appears to be order-disorder based on Mn K edge X-ray absorption spectroscopy~\cite{Yamaguchi1998Jahn-TellerSpectroscopy}. 
Electron spin resonance measurements on CuSb$_2$O$_6$ indicate persistence of local JT distortions through the JT transition~\cite{Heinrich2003StructuralESR} suggesting an order-disorder transition. 

cJT distortions are generally suppressed due to the additional thermal energy in the system~\cite{Gehring1975Co-operativeEffects}.
Expansion of the crystal lattice weakens the effective interaction between local JT centers mediating cooperativity, reducing the enthalpy gain associated with orbital ordering. 
Increasing thermal energy also enables octahedra to dynamically cycle through axes of elongation corresponding to different local energy minima, which would suppress a time-averaged cJT distortion \textit{via} a dynamic JT fluctuation~\cite{Sicolo2020AndSystem}. 
Additionally, free energy becomes dominated by entropic contributions over enthalpic contributions on heating, and so enthalpy minimisation due to cJT will be less of a factor in determining the crystal structure~\cite{Ahmed2005TheModel}; this is important as there may be significant configurational entropy associated with orbital disorder depending on the geometry of the lattice in question~\cite{Nagle-Cocco2024DisplaciveNaNiO2}. 
The common interpretation of JT transitions as order-disorder can therefore be explained in terms of maximising configurational entropy at high temperature. 

\begin{figure*}[t]
    \includegraphics[scale=1]{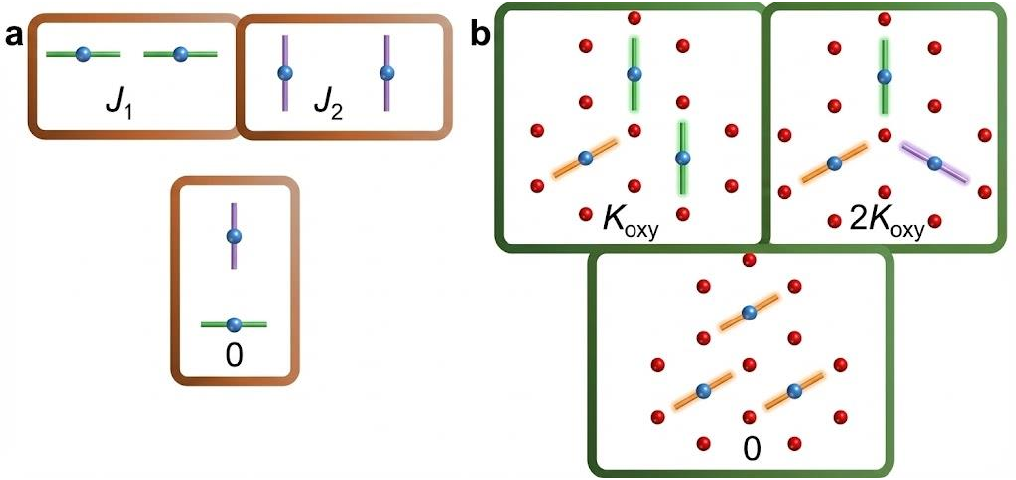}
    \captionsetup{list=no} 
    \caption{
        \label{geometry-term.pdf}
        Diagram showing the JT configurations associated with zero and non-zero geometric energies as used in Equations~\ref{Potts_model_modified_perovskite} and \ref{geometry_term_nickelate}. 
        (a) The case for perovskites, which is a reproduction of the Potts model of Ahmed and Gehring~\cite{Ahmed2005TheModel,Ahmed2006PottsLaMnO3,Ahmed2009VolumeModel}. 
        (b) The case for intra-layer nearest-neighbour NiO$_6$ octahedra within layered nickelates assuming all sites have $\rho=1$. 
        Blue solid circles indicate JT-active sites, red solid circles (for the layered nickelate case) indicate O atoms, and solid lines indicate JT elongations.
    }
\end{figure*}

The nature of the JT transition in the previously-discussed layered nickelate materials has attracted much interest~\cite{Nagle-Cocco2022PressureNaNiO2,Nagle-Cocco2024DisplaciveNaNiO2,Radin2020Order-disorderMaterials,Nagle-Cocco2025Dome-likeNaNiO2,Genreith-Schriever2024Jahn-TellerDiffraction,Phillips2025Collinearsub2/sub}. 
Computational work based on Monte Carlo simulations integrated with density functional theory (DFT)~\cite{Radin2020Order-disorderMaterials} concluded that although the parameter phase space for layered nickelates can accommodate both an order-disorder and a displacive transition, an order-disorder transition is more likely in common with previously-studied JT transitions.~\cite{Radin2020Order-disorderMaterials} 
However, more recently, X-ray absorption spectroscopy~\cite{Nagle-Cocco2024DisplaciveNaNiO2,Jacquet2025Temperature-Dependentsub2/sub}, neutron total scattering~\cite{Nagle-Cocco2024DisplaciveNaNiO2}, and \textit{ab initio} molecular dynamics simulations~\cite{Nagle-Cocco2024DisplaciveNaNiO2,Genreith-Schriever2024Jahn-TellerDiffraction} have indicated that layered nickelates exceptionally exhibit displacive JT transitions, in which there is no minimum in the free energy landscape associated with JT distortions above the transition temperature. 
This was attributed~\cite{Nagle-Cocco2024DisplaciveNaNiO2} to the subextensive configurational entropy associated with orbital disorder on a 2D triangular lattice in which the axes of elongation are geometrically constrained by the chemical instability associated with having multiple axes of elongation pointing at a common O anion. 
In contrast, we note that evidence for an order-disorder transition in non-stoichiometric LiNiO$_2$ with site-mixing has also been reported from X-ray absorption spectroscopy measurements~\cite{Jacquet2025Temperature-Dependentsub2/sub}. 

While the observation of displacive behaviour in nickelates appeared exceptional at the time they were reported, a recent study~\cite{Batnaran2025TheLaMnO_3} has shown that experimental total scattering data for the LaMnO$_3$ case can be reproduced without well-defined and prevalent octahedral elongations. 
This would indicate that dynamics, rather than configurational entropy, is the key driver of behaviour through JT transitions. 
This raises the possibility that displacive JT transitions are more common than previously assumed, and that many historical assignments of order-disorder character may warrant re-examination. 

The cJT distortion in perovskites was previously studied using Monte Carlo simulations by Ahmed and Gehring (2005)~\cite{Ahmed2005TheModel} and applied to the case of LaMnO$_3$~\cite{Ahmed2006PottsLaMnO3,Ahmed2009VolumeModel}. In their model, an anisotropic Potts variable on each site successfully reproduced the cJT ordering at room-temperature in LaMnO$_3$~\cite{Ahmed2005TheModel,Ahmed2006PottsLaMnO3} and the dependence of lattice parameters on temperature through the JT transition~\cite{Ahmed2009VolumeModel}. 
However, their simple model did not include amplitudes of JT distortion or the possibility of JT-undistorted MnO$_6$ octahedra, and so an order-disorder transition was implicitly built into the model. 
More recently, Tragheim \textit{et al.} (2025)~\cite{tragheim2025interplay} developed a more complex model in which JT amplitude may vary as a consequence of the octahedral tilts and shapes, which are the varying parameters; however, they did not apply this directly to the temperature-induced JT transition. 

For the layered nickelates, Radin \textit{et al.}~\cite{Radin2020Order-disorderMaterials} used an anharmonic vibrational Hamiltonian with continuous JT modes and Monte Carlo sampling, parametrized from DFT. 
Unlike the model of Ahmed and Gehring, this more complex approach~\cite{Radin2020Order-disorderMaterials} contains a regime in the phase space of starting parameters, termed the ``stiff-lattice regime", which can lead to displacive behaviour. 
However, they predicted that layered nickelates would occur in the soft-lattice regime where order-disorder transitions are predicted by the Monte Carlo simulations. 

In this present work, we extend the relatively simple model of Ahmed and Gehring with variable amplitude, $\rho$, of JT distortion on each site and a single-ion term in the Hamiltonian to simulate the enthalpy saving associated with JT distortion. 
Using this model, we simulate the JT transitions in perovskites such as LaMnO$_3$. We then develop a similar, simple model for the layered nickelates. 

We find that the mean magnitude of JT distortion decreases through the transition but remains non-zero, with the high-temperature magnitude being larger in the perovskite lattice than the 2D nickelate lattice, consistent with experimental results which are interpreted in terms of an order-disorder transition. 
However, the distribution of values of JT magnitude does not resemble that at low temperatures. 
For an order-disorder transition, we would expect to see a well-defined peak in the distribution at finite, non-zero $\rho$, with negligible density at $\rho=0$. 
Instead, we see little preference for any particular $\rho$ value at high temperatures for perovskites, which appears consistent with dynamic exploration of the $E_g$($Q_2$,$Q_3$) phase space. 
\textcolor{black}{
Our findings therefore suggest that a Potts-like model with variable JT amplitude is consistent with the interpretation of Batnaran et al.~\cite{Batnaran2025TheLaMnO_3}, in which the high-temperature JT phase departs from the conventional order--disorder picture of persistent local distortions. 
Our model provides a minimal framework for exploring departures from the fixed-amplitude order–disorder limit.
}

\section{Methodology}

\subsection{Hamiltonian}

We define a global classical Hamiltonian, for a system with \textit{N} JT-active octahedra, as follows:

\begin{equation}\label{hamiltonian_equation}
    H = E_\mathrm{single-ion} 
    + E_\mathrm{geometry}
\end{equation}

\subsubsection{Single-ion term}

In this Hamiltonian, $E_\mathrm{single-ion}$ is the enthalpic impetus to JT distortion. It is defined following a Landau-like second-order expansion of the free energy~\cite{Gehring1975Co-operativeEffects} thus:

\begin{equation}\label{E_single-ion_term}
    E_\mathrm{single-ion} = \sum_i^N \left[ 
        \alpha \rho_i^2 + \beta \rho_i^4 
    \right]
\end{equation}

in which $\rho_i$ is the magnitude of the JT distortion at the \textit{i}th JT-active site. 
$\alpha$ and $\beta$ are empirical parameters which determine the functional dependence of enthalpy on $\rho$. 
Typically, we would expect $\alpha<0$, $\beta>0$, and $|\beta| < |\alpha|$. 
In the simulations presented here, where variable-magnitude $\rho$ is enabled, \textcolor{black}{$\alpha=-1$ and $\beta=1/2$, with the ratio $\beta=-\alpha/2$ chosen arbitrarily so that $E_\mathrm{single-ion}$ is minimized when $\rho=1$}. 
\textcolor{black}{The magnitudes of $\alpha$ and $\beta$ places $E_\mathrm{single-ion}$ approximately on the energy scale as $E_\mathrm{geometric}$. In SI Section~\ref{SI_Section_varying_alpha_magnitude} we show that the results are qualitatively unchanged if we increase or decrease the value of $\alpha$ relative to $E_\mathrm{geometric}$.}

\begin{figure*}[t]
    \includegraphics[scale=1]{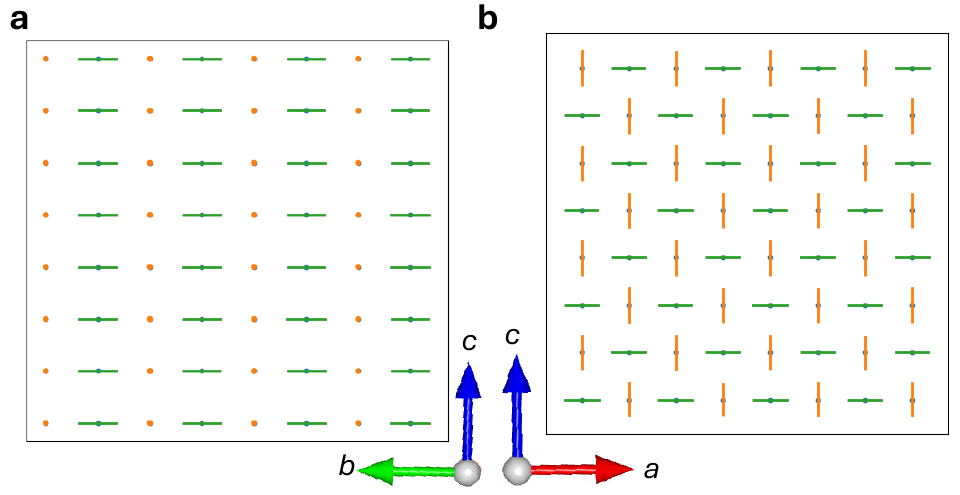}
    \captionsetup{list=no} 
    \caption{
        \label{perovskite_C-type_lowT_simulation.pdf}
        The results of a Monte Carlo \textcolor{black}{simulation} with dynamic $\rho$ ($P_\mathrm{switch}=1/2$), reproducing Phase 5 (C-type) of the Ahmed and Gehring~\cite{Ahmed2005TheModel} phase diagram for the anisotropic Potts model in an $8\times8\times8$ perovskite supercell. 
        (a) Example configuration in the $bc$-plane in a randomly-selected cross-section. 
        (b) Example configuration in the $ac$-plane in a randomly-selected cross-section. 
        Energy parameters $\alpha=-1$,$\beta=1/2$ from equation~\ref{E_single-ion_term}. 
        Simulated annealing was used before settling on a final $T=0.001$. 
        This run lasted for $10^7$ iterations. 
        Plots of energy and $\langle \rho\rangle$ with iteration can be seen in Figure~\ref{figure_dynamic-rho_phase5}. 
    }
\end{figure*}

\subsubsection{Geometry term}

The $E_\mathrm{geometry}$ term is defined uniquely based on the structure in question. 
For the case of perovskites, this $E_\mathrm{geometry}$ term is equivalent to the anisotropic Potts model Hamiltonian defined by Ahmed and Gehring~\cite{Ahmed2005TheModel,Ahmed2006PottsLaMnO3,Ahmed2009VolumeModel}, reproduced as follows over all nearest-neighbour JT-active pairs $i,j$:

\begin{equation}\label{Potts_model_modified_perovskite}
    E_\mathrm{geometry}^\mathrm{perovskite}
    = \frac{1}{2} \sum_{\langle i,j \rangle}
    w(\rho_i,\rho_j) \,
    J_{S_i}(\hat{\mathbf{e}}_{ij}) \,
    \delta_{S_i,S_j},
\end{equation}

In this equation, $S_i$ is the direction of octahedral elongation at JT-active B-site cation $i$, representing the orbital state of the JT-active B-site cation; it can take one of three possible values. 
$\delta_{S_i, S_j}$ is the Kronecker delta which enforces that only like-orbital pairs contribute to the interaction energy. 
$\hat{\mathbf{e}}_{ij}$ is the unit vector pointing from site $i$ toward site $j$.
$J_{S_i}(\hat{\mathbf{e}}_{ij})$ is the anisotropic interaction constant, which takes a value $J_1$ if the direction of $S_i=S_j$ matches $\hat{\mathbf{e}}_{ij}$, a value of $J_2$ if $S_i=S_j$ is perpendicular to $\hat{\mathbf{e}}_{ij}$, or a value of zero otherwise, as shown in Figure~\ref{geometry-term.pdf}(a). 
The prefactor $\tfrac{1}{2}$ avoids double counting of site pairs in the nearest-neighbor sum $\langle i,j \rangle$. 

The multiplier $w(\rho_i,\rho_j)$ is used in this case to reduce the contribution to the geometry term of undistorted octahedra without incentivising infinite distortion. It is defined thus:

\begin{equation}
   w(\rho_i,\rho_j)  = 2\rho_i \rho_j/(1+\rho_i \rho_j)
\end{equation}

The possible values of $J_{S_i}(\hat{\mathbf{e}}_{ij})$, $J_1$ and $J_2$, are empirical constants, the values of which will determine the ground-state cJT ordering. 
Ahmed and Gehring found that~\cite{Ahmed2006PottsLaMnO3} the cJT in room-temperature LaMnO$_3$ is reproduced by $J_1>0$, $J_2<0$, and $|J_2/J_1| < 1/2$. 
The collinear JT ordering found, for instance, in Ba$_{0.5}$La$_{0.5}$CoO$_3$~\cite{Fauth2002IntermediateDistortions} could be reproduced for $J_2<0$ and $J_1<-J_2$. 

For the layered nickelates, the geometric energy term is based on the chemical instability associated with under-bonding of O anions. 
For instance, an earlier work~\cite{Chung2005LocalDiffraction} proposed a trimer-like cJT ordering in LiNiO$_2$ in which three elongated Ni-O bonds point at the same oxygen anion, but subsequent computational studies using DFT~\cite{Chen2011First-principleLiNiO2} showed this to be highly unstable relative to collinear or zigzag models in which a single elongated bond points at each oxygen anion. 
We calculate this by summing over all O anions in the simulation as follows: 

\begin{equation}\label{geometry_term_nickelate}
E_{\mathrm{geometry}}^\mathrm{nickelate}
= \frac{K_{\mathrm{oxy}}}{2} \sum_{O}
\left[
\max\!\bigl(0,\; s_O - 1 \bigr)
\right]^2 
\end{equation}

where $K_{\mathrm{oxy}}$ is an empirical parameter, $K_{\mathrm{oxy}}>0$, which determines the strength of this O under-bonding penalty, with the implication of this equation shown in Figure~\ref{geometry-term.pdf}(b) assuming all $\rho$ are 1. 
The ``underbonding density" term $s_O$ is given by summing over all three nearest-neighbour Ni cations to a given O anion thus:

\begin{equation}
s_O \;=\; \sum_{i \in N(O)} \; \delta_{\vec{\rho}_i , \,\vec{r}_{iO}} \; \rho_i ,
\end{equation}

where $\delta_{\vec{\rho}_i , \,\vec{r}_{iO}}$ is 0 or 1 depending on whether the elongated Ni-O bond ($\vec{\rho_i}$) matches the vector between the Ni and the O anion ($\vec{r}_{iO}$). 

We note that the nature of the Hamiltonian we use here renders us insensitive to the possibility of disproportionation as proposed in some prior works for both layered nickelates~\cite{Chen2011ChargeStudy,Foyevtsova2019LiNiO2Glass,Poletayev2025Temperature-dependentLiNiO2} and perovskites~\cite{garcia1994neutron,mizokawa2000spin,garcia2009structure,pascut2023role}. 

The perovskite and nickelate $E_\mathrm{geometry}$ term share the same degrees of freedom (a three-state elongation axis and a continuous amplitude $\rho$), but differ qualitatively in their interaction terms: pairwise orbital alignment in the perovskite case, versus oxygen-underbonding constraints in the nickelate case. 
Consequently, the nickelate geometry term is no longer a Potts Hamiltonian as the perovskite geometry term is. 

\subsection{Model cell}

Simulations are run over periodic cells in two- or three-dimensions for the layered nickelate and perovskite case, respectively. 
For the perovskite case, a cubic primitive cell is used with an $8\times8\times8$ grid of metal cations (512 atoms). 
For the layered nickelate case, the cell is a $10\times30$ rhombus-shaped grid of Ni cations (300 atoms), where the two basis vectors enclose 60$^\circ$. 
The use of a two-dimensional lattice for the layered nickelates relies on the axiom that inter-layer interactions are not significant for the JT ordering. 
Unless otherwise indicated, the starting configuration for all runs was one in which JT distortions were randomly distributed within the unit cell, all with identical magnitude $\rho=1$.

\subsection{Metropolis algorithm}

The Monte Carlo simulation runs over a number of iterations ($10^5$ to $10^7$) using the Metropolis algorithm~\cite{Metropolis1953EquationMachines}. 
At each iteration, a metal site is selected at random. 
With probability $0 \leq P_\mathrm{switch} \leq 1$, a $\rho$ shift occurs in which the magnitude of $\rho$ is varied randomly by some amount, with $-0.1 \leq \Delta \rho \leq 0.1$ (with $\rho$ capped at 0 at the lower limit and 1.5 at the upper limit). 
Alternatively, with probability $P_\mathrm{flip}=1-P_\mathrm{switch}$, a direction shift occurs, in which the axis of elongation of the metal site is shifted randomly. 
Either way, energy is then re-calculated according to Equation~\ref{hamiltonian_equation}. 
If $\Delta E \leq 0$, the change is accepted. 
Otherwise, the change is accepted with probability $\exp \left[-\Delta E/T \right]$. 
For most testing in this work, $P_\mathrm{switch}$ was set arbitrarily at 0.5. 

Where indicated, a simulated annealing procedure is used to avoid being stuck in local minima. This is discussed in Appendix A.

\section{Results: JT-distorted perovskite}

\begin{figure}[t]
    \includegraphics[scale=0.65,trim={0.25cm 0.25cm 0.25cm 0.25cm},clip]{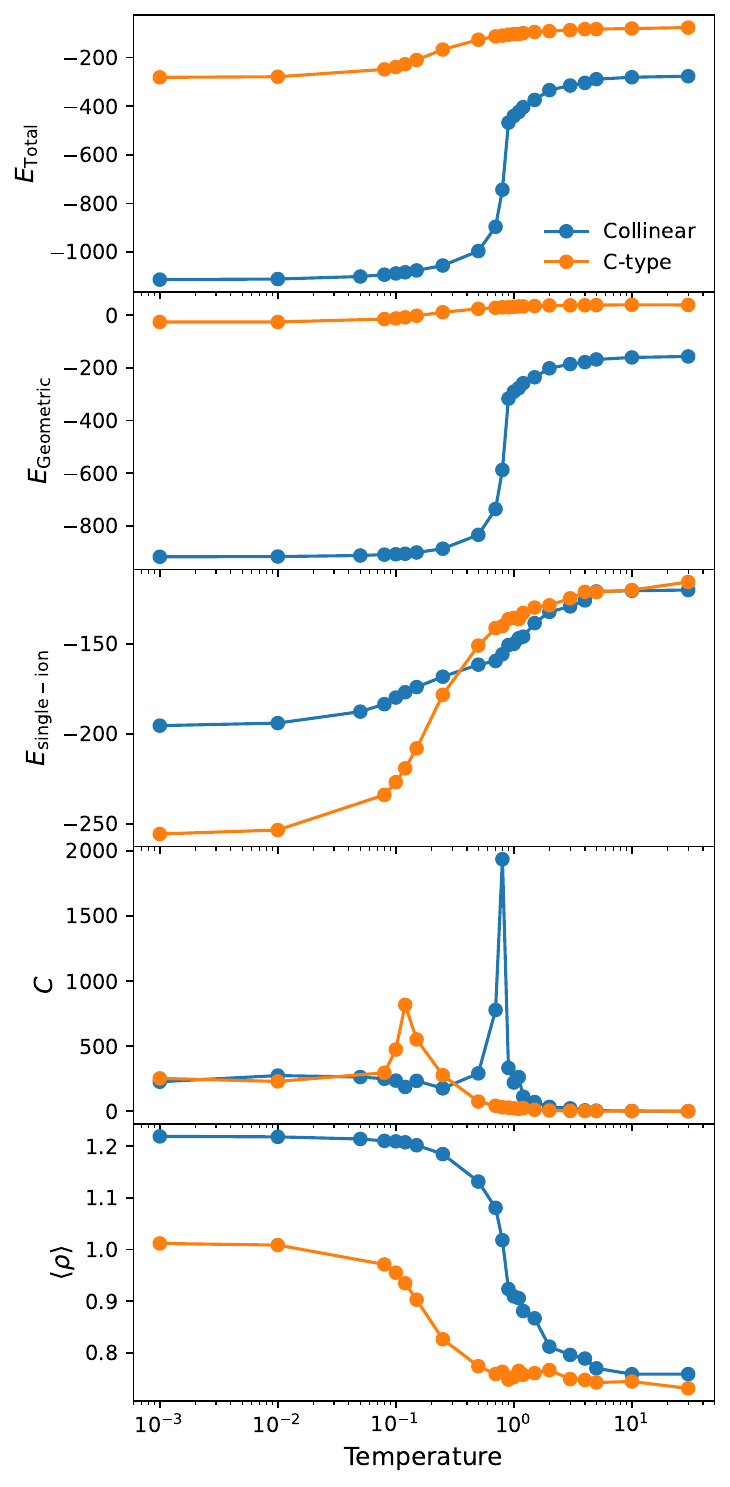}
    \captionsetup{list=no} 
    \caption{
        \label{perovskite_VT_simulation.pdf}
        The mean energy terms $E_\mathrm{total}$, $E_\mathrm{geometric}$ (Eq~\ref{Potts_model_modified_perovskite}), and $E_\mathrm{single-ion}$ (Eq~\ref{E_single-ion_term}), heat capacity $C$, and mean $\langle\rho\rangle$, averaged over the final 10\% of iterations in Monte Carlo simulations, as a function of temperature for the collinear and C-type orbital orderings in an $8\times8\times8$ perovskite lattice. 
        The Monte Carlo simulations ran for $10^7$ iterations in total at each temperature. 
        This figure is reproduced on a non-logarithmic temperature scale in SI Figure~\ref{perovskite_VT_simulation_nonlog}.
    }
\end{figure}

Before allowing $\rho$ to vary (i.e. by setting all metal sites to have constant non-zero JT magnitude, and setting $P_\mathrm{switch}=0$), we tested our model where the Hamiltonian consisted of the $E_\mathrm{geometry}^\mathrm{perovskite}$ term only (i.e. $\alpha=\beta=0$), as shown in SI Section~\ref{J1_J2_reproduction_section}. 
This reproduced all phases of the $(J_1,J_2)$ anisotropic Potts model phase diagram of Ahmed and Gehring~\cite{Ahmed2005TheModel}, validating their results and our simulation approach. 

\begin{figure*}[t]
    \includegraphics[scale=1.0]{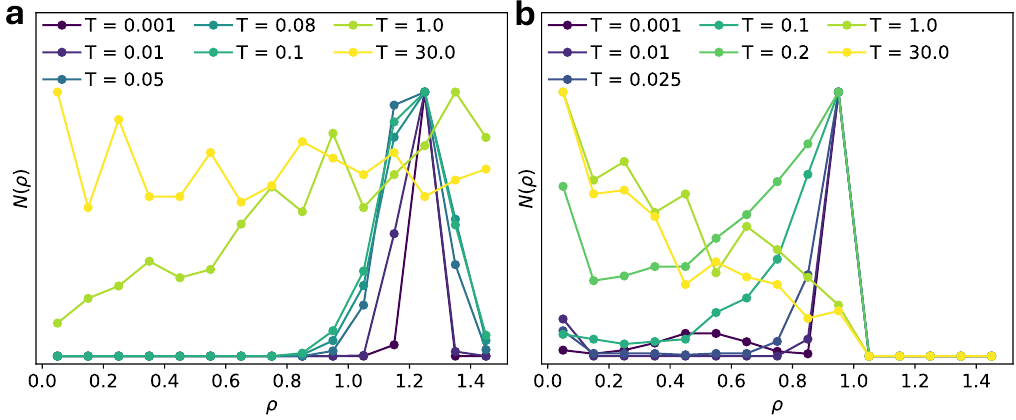}
    \captionsetup{list=no} 
    \caption{
        \label{nickelate-histograms.pdf}
        Histograms of $\rho$ for (a) $8\times8\times8$ perovskite collinear ordered supercell and (b) the final $10\times30$ nickelate layer with $K_\mathrm{oxy}=10^{10}$ after $10^7$ iterations, with simulated annealing. 
        For other configurations see Figures~\ref{nickelate_Koxy1_rho_histogram.pdf} to \ref{phase5_rho_histogram.pdf}.
    }
\end{figure*}

\begin{figure}[t]
    \includegraphics[scale=0.55,trim={0.25cm 0cm 0cm 0cm},clip]{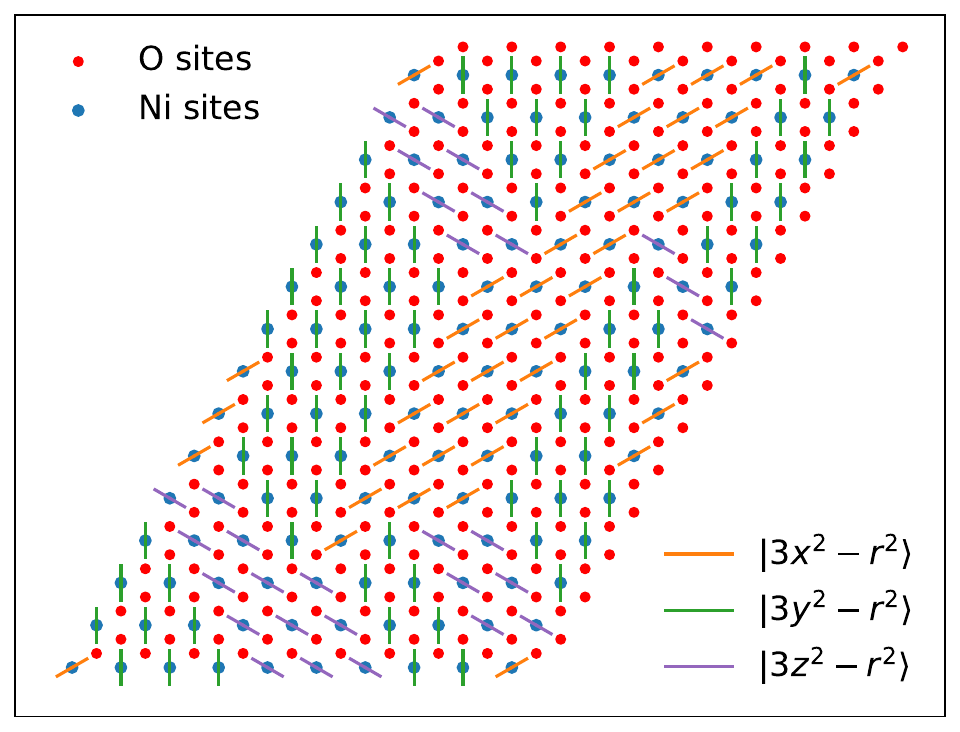}
    \captionsetup{list=no} 
    \caption{
        \label{JT_ordered_lowT_nickelate.pdf}
        An example converged configuration of a $10\times15$ layered nickelate. 
        This simulation used a static $\rho$ ($P_\mathrm{switch}=0$; $\alpha=\beta=0$) model, with $K_\mathrm{oxy}=10^{10}$ and $T=0.001$ after simulated annealing (with $T_\mathrm{max}=10$), and was run for $10^6$ iterations. 
        Collinear domains can be seen, often bound by defects of over-bonded (no JT axes) or under-bonded (multiple JT axes) oxygen anions, indicating that this is a local, but not global, energy minimum. 
        In this simulation, we set $K_\mathrm{Oxy}=10^{10}$. 
    }
\end{figure}

Two phases of the Ahmed and Gehring phase diagram are of particular interest which we call ``collinear" and ``C-type" (phase 1 and phase 5 using Ahmed and Gehring's notation). Collinear is a ferro-orbital model with parallel axes of elongation, and C-type consists of alternating stacked elongations within a single plane [Figure~\ref{JT-ordering-from-literature}(a)]. 
These are of interest because they have been reported experimentally. 
We henceforth only study these two phases for perovskites. 
The next step was to test whether, for $T<<1$, adding a dynamic JT magnitude with the $E_\mathrm{single-ion}$ term enables the recovery of these two phases. 

\textcolor{black}{Figure~\ref{perovskite_C-type_lowT_simulation.pdf} shows cross-sections of the final perovskite supercell for the ``C-type" ordering (where $J_1=1$ and $J_2=-0.1$), indicating successful convergence with the expected orbital order; energy and mean JT magnitude $\langle\rho\rangle$ with iteration are shown in Figure~\ref{figure_dynamic-rho_phase5}.} 
Figure~\ref{figure_dynamic-rho_phase1} shows the corresponding information for the collinear case. In both cases, the dynamic $\rho$ model is able to reproduce the phase diagram given by the static $\rho$ model~\cite{Ahmed2005TheModel}. 

We next consider the variable-temperature behaviour of these orbital orderings. We performed simulations, with simulated annealing, at a series of temperatures up to $T=30$, for both the collinear and C-type ($J_1$,$J_2$) configurations. 
Figure~\ref{perovskite_VT_simulation.pdf} shows the mean energy and $\langle\rho\rangle$ for the final 10\% of iterations in the simulations as a function of temperature, for both cases. 
We see that, as temperature increases, there is a transition, which can be seen best in the heat capacity data [Figure~\ref{perovskite_VT_simulation.pdf}], where heat capacity is the variance in $E_\mathrm{total}$ divided by $T^2$. 
$\langle\rho\rangle$ decreases with heating, with the rate of decrease being low on approach to the transition, and then a rapid decrease before settling at around $\langle\rho\rangle\approx 0.8$. 
We have tested the resilience of this finding against varying cell size [Figures~\ref{SI_size-with-E_perov_4x4x4.pdf} and \ref{SI_size-with-E_perov_6x6x6.pdf}] with little variation. 
\textcolor{black}{We use representative points in the centre of the collinear and C-type phases in ($J_1$,$J_2$) space, but in SI Section~\ref{SI_section_check_boundaries_of_phases_VT_behaviour} we show that our findings are qualitatively consistent throughout these regions of phase space}. 
The finding that $\langle\rho\rangle$ decreases on heating even in the low-temperature regime approximately corresponds to experimental observations from diffraction on LaMnO$_3$ that $d\rho/dT<0$, although KCuF$_3$ exhibits the opposite trend~\cite{Marshall2013UnusualDiffraction}. 

A convergence of $\langle\rho\rangle\rightarrow0.8$ at high temperatures appears consistent with the experimental reports (from PDF and EXAFS) of JT-like octahedral distortion at high temperatures, if we interpret it as the system remaining in the enthalpic minimum imposed by Eq~\ref{E_single-ion_term}. 
However, when we plot histograms of $\rho$ in the final configuration [Figure~\ref{nickelate-histograms.pdf}], we see that the distribution at high temperature does not resemble the distribution at low temperature, which would be the signature of an order-disorder transition. 
Instead, we see $N(\rho)$ appears invariant with the magnitude, which suggests that rather than a persistence of JT distortions the system is merely dynamically exploring the available phase space of possible values of $\rho$, which is consistent with the recent theoretical study of LaMnO$_3$~\cite{Batnaran2025TheLaMnO_3}. 
These findings could be interpreted more in terms of a displacive transition than an order-disorder transition, although the invariance in $N(\rho)$ with $\rho$ indicates that if there is a free energy minimum at the origin in $E_g(Q_2,Q_3)$ phase space, it is not significantly more stable than other regions of the phase space.

\section{Results: Layered nickelate}

\begin{figure}[ht]
    \includegraphics[scale=0.65,trim={0.25cm 0.3cm 0.25cm 0.25cm},clip]{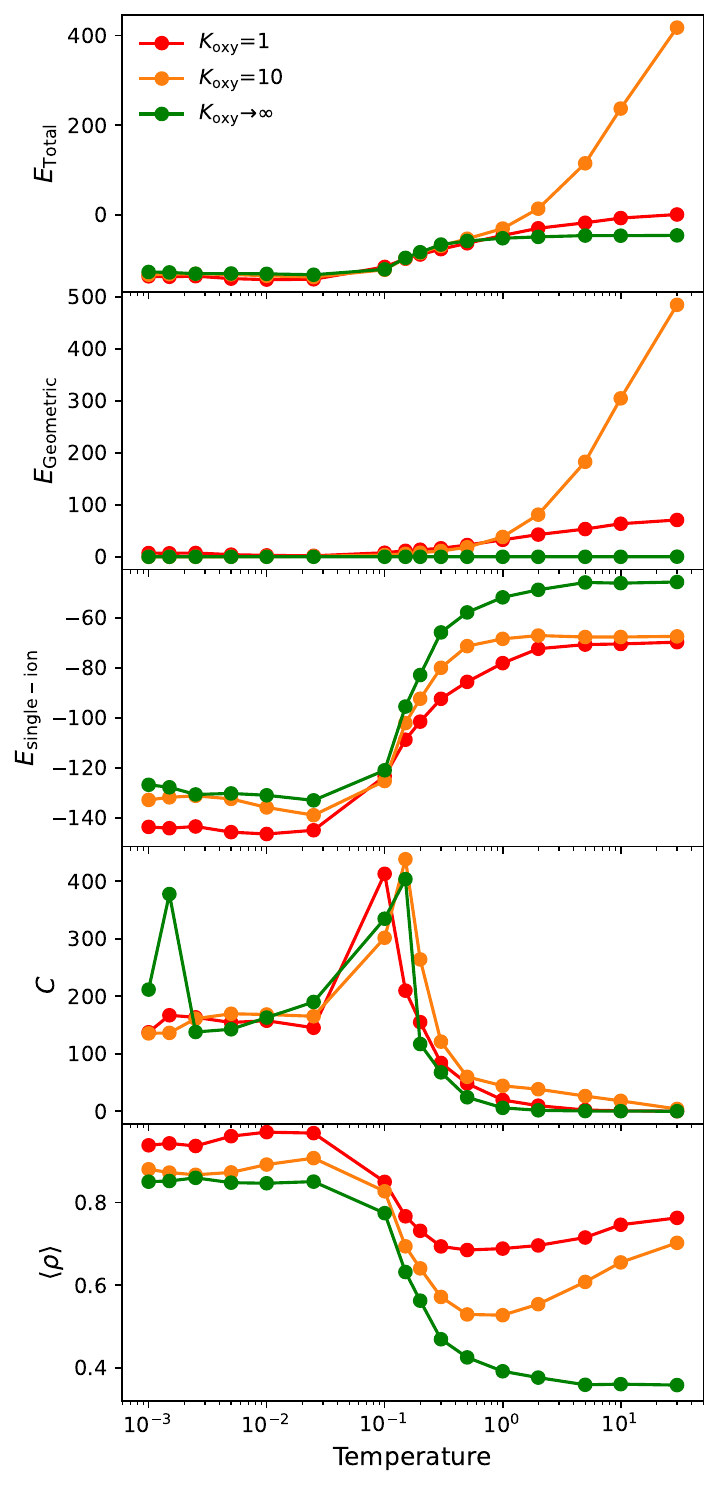}
    \captionsetup{list=no} 
    \caption{
        \label{nickelate_VT_simulation.pdf}
        The mean energy terms $E_\mathrm{total}$, $E_\mathrm{geometric}$ (Eq~\ref{geometry_term_nickelate}), and $E_\mathrm{single-ion}$ (Eq~\ref{E_single-ion_term}), heat capacity $C$, and mean $\langle\rho\rangle$, averaged over the final 10\% of iterations in Monte Carlo simulations, as a function of temperature for the $10\times30$ nickelate lattice. 
        The Monte Carlo simulations ran for $10^7$ iterations in total at each temperature. 
        We present these results as a function of the strength of the oxygen under-bonding penalty $K_\mathrm{oxy}$.
        This is plotted on a non-logarithmic temperature scale in Figure~\ref{nickelate_VT_simulation_not-log}. 
    }
\end{figure}

\begin{figure}[htb]
    \includegraphics[scale=0.6]{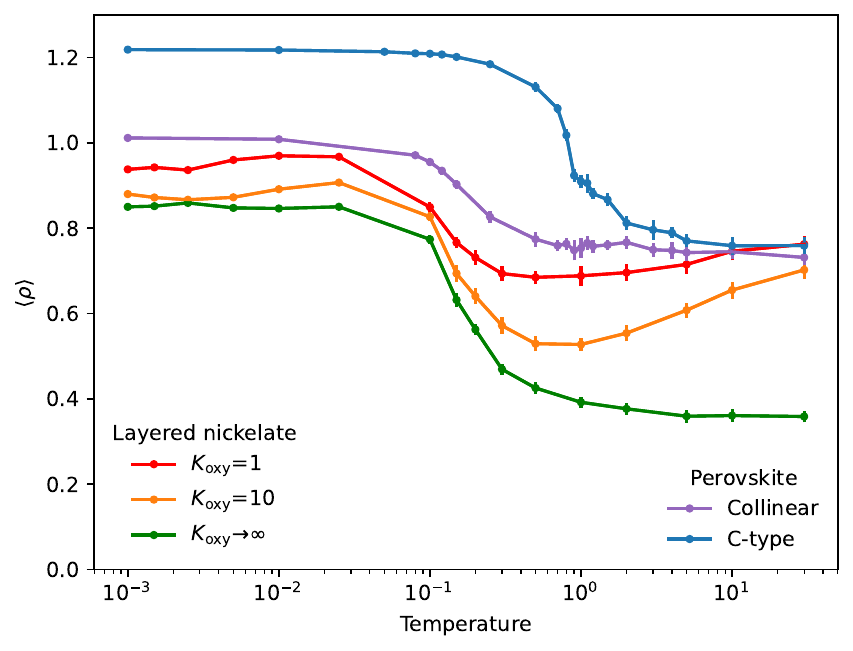}
    \captionsetup{list=no} 
    \caption{
        \label{JT-magnitude-with-temp.pdf}
        $\langle\rho\rangle$ with temperature, comparing phase 1 (collinear) and phase 5 (C-type) perovskite ordering with layered nickelates for various $K_\mathrm{oxy}$. 
        The corresponding figure with non-logarithmic temperature scaling is Figure~\ref{JT-magnitude-with-temp_non-log}. 
    }
\end{figure}

We next simulated the layered nickelates. 
Here, we rely on the axiom that inter-layer effects are negligible due to the alternating \textit{A}O$_6$ octahedral layers ($A$=Li,Na), and consider only single layers. 
As with the perovskites, we first performed Monte Carlo simulations for the static $\rho$ case where all sites begin JT-distorted, with $P_\mathrm{switch}=0$, and the Hamiltonian consisting solely of the $E_\mathrm{geometry}^\mathrm{nickelate}$ term (i.e. $\alpha=\beta=0$), with $T<<1$. 
An example of a converged layer configuration is shown in Figure~\ref{JT_ordered_lowT_nickelate.pdf}. 
We see the emergence of ordered collinear domains, as reported from diffraction studies on NaNiO$_2$~\cite{Dyer1954AlkaliMNiO2,Dick1997TheScattering,Sofin2005NewNaNiO2,Nagle-Cocco2022PressureNaNiO2} and stoichiometric LiNiO$_2$~\cite{Phillips2025Collinearsub2/sub}. 
The boundary between these domains resembles the ``zigzag" ordering which has been proposed as a possible ground state for LiNiO$_2$ on the basis of DFT studies~\cite{Chen2011First-principleLiNiO2,Foyevtsova2019LiNiO2Glass}. 
Defects then occur when two or three domains intersect. 
Such defects involve either over-bonded (no JT axes) or under-bonded (multiple JT axes) oxygen anions, and resemble the dimer or even trimer models proposed by Chung \textit{et al} (2005)~\cite{Chung2005LocalDiffraction} [Figure~\ref{SI_LNO-ordering-from-lit.pdf}(e,h)] which have been found from DFT to be highly unstable~\cite{Chen2011First-principleLiNiO2}. 
This domain effect, in contrast, does not seem to be important for the perovskites, likely because they can relieve frustration by stacking patterns along $c$, which reduces the likelihood of frustrated interactions. 

For $T<<1$, we see this behaviour is preserved when working with the dynamic $\rho$ and non-zero $E_\mathrm{single-ion}$ term, with $K_\mathrm{oxy}$ arbitrarily large (i.e. $10^{10}$).  
From here, we iteratively increase temperature and evaluate the changes in $\langle\rho\rangle$, as shown in Figure~\ref{nickelate_VT_simulation.pdf}. As with the perovskites, the heat capacity $C$ shows that an ordering transition occurs, above which there is local disorder. 
We see that, keeping $K_\mathrm{oxy}$ arbitrarily large, as $T\rightarrow\infty$, $\langle\rho\rangle\rightarrow0.4$, much lower than the perovskite case. 
Figure~\ref{nickelate-histograms.pdf} shows histograms of $\rho$ at low and high temperatures, with most Ni sites exhibiting a large JT distortion at low-$T$ (except a few Ni sites which occur at the boundary between domains), and Ni sites at high temperatures exhibiting no preference for JT distortion, with the highest distribution of sites occurring at low-$\rho$. 
This suggests that the preference for large $\rho$ and JT distortions vanishes on the 2D nickelate lattice, which is displacive-like behaviour. 
Although our simulations for the perovskites were broadly consistent with a similar conclusion, there clearly exist differences between these two cases: (1) the significantly reduced $\langle\rho\rangle$ at high $T$ for the nickelate case compared with the perovskite case and (2) the decreasing $N(\rho)$ with $\rho$ for the nickelate case compared with the approximate invariance of $N(\rho)$ with $\rho$ in the perovskite case. 

In Ref.~\cite{Nagle-Cocco2024DisplaciveNaNiO2}, we presented a mathematical proof that JT disorder in layered nickelates has subextensive configurational entropy, and argued that this drives the displacive behaviour in layered nickelates. 
One of the axioms of this proof is that oxygen underbonding is unfeasible and so multiple axes of JT elongation do not point at the same oxygen; this is equivalent to our choice in these Monte Carlo simulations to set $K_\mathrm{oxy}\rightarrow\infty$. 
In Figure~\ref{nickelate_VT_simulation.pdf}, we also show the temperature-dependence of $\langle\rho\rangle$ if we reduce the magnitude of $K_\mathrm{oxy}$ (while retaining $K_\mathrm{oxy} > 0$), i.e. resulting in scenarios such as the trimer or dimer JT-ordering being energetically unfavourable but not totally unfeasible as for the $K_\mathrm{oxy}\rightarrow\infty$ case. 
\textcolor{black}{
This results in $\langle\rho\rangle$ converging much higher, similar to the case for the perovskites, and with distributions of $\rho$ exhibiting no preference for particular $\rho$ values [Figure~\ref{nickelate_Koxy1_rho_histogram.pdf}]. 
Despite the simplicity of the model, this suggests that the oxygen-underbonding penalty can strongly influence the high-temperature $\rho$ distribution by restricting the configurational entropy of orbitally-disordered states. Within this framework, lattice geometry therefore provides a mechanism for producing different high-temperature JT behaviour in different model systems.
}

\section{Conclusion}

In this study, we successfully extended the Monte Carlo model of Ahmed and Gehring~\cite{Ahmed2005TheModel,Ahmed2006PottsLaMnO3,Ahmed2009VolumeModel} to include variable Jahn--Teller (JT) distortion amplitudes, enabling us to retain the simplicity of the Ahmed and Gehring model, while allowing for the possibility that JT distortions may vanish locally at high temperature. 
The results of this approach are qualitatively consistent with both experimental evidence for persistence of JT-like distortion modes above JT transitions in perovskites~\cite{Qiu2005OrbitalLaMnO3,Chatterji2003VolumeTransition} (which, in our simulations, manifests as a high $\langle\rho\rangle$), and recent computational results which suggest that the high-temperature state of JT-distorted perovskites is not characterised by well-defined JT-distorted octahedra~\cite{Batnaran2025TheLaMnO_3}, but rather a dynamic sampling of the ($Q_2$,$Q_3$) phase space. 
\textcolor{black}{More generally, these results show that allowing amplitude fluctuations qualitatively alter the high-temperature behaviour of Potts-like JT models.} 

\textcolor{black}{Despite our finding that our variable-amplitude model does not show persistent JT distortions at high temperature,} there are clearly differences between the two systems due to the geometry of the lattice. 
We reported that the converged $\langle\rho\rangle$ of the perovskite lattices is much higher ($\sim$0.8) than the nickelate lattice ($\sim$0.4) at high temperature, and as shown in Figure~\ref{nickelate-histograms.pdf} the number distribution of distortion magnitudes appears invariant with $\rho$ for the perovskite case but decreases as $\rho$ increases for the nickelate case. 
This is likely a consequence of the subextensivity of configurational entropy of disordered octahedral elongations on the nickelate lattice, as previously reported~\cite{Nagle-Cocco2024DisplaciveNaNiO2}. 
\textcolor{black}{If this were to occur in real systems, it would likely} manifest experimentally in a larger skewness for the first metal-oxygen peak in the experimental pair distribution function for perovskites like LaMnO$_3$, as compared with nickelates like NaNiO$_2$. 
This may be a factor in the historic attribution of order-disorder behaviour to the JT transitions in the perovskites. 

Besides the layered nickelates and perovskites studied in this work, the model presented here may be applied to other JT-distorted systems of interest. 
For instance, spinels, double perovskites, and Prussian Blue analogs all exhibit cJT distortions. 
Future work should investigate these systems, and benchmark findings against local probe experiments such as total scattering and EXAFS. 
Additionally, useful future work would apply the more complex anharmonic vibrational Hamiltonian of Radin \textit{et al.}~\cite{Radin2020Order-disorderMaterials} to perovskite systems such as LaMnO$_3$ to better understand the recent developments in our understanding of these systems. 

\section*{Software}

Code was written in Python 3~\cite{VanRossum1995PythonTutorial}, with assistance from generative large-language models ChatGPT~\cite{OpenAI2024GPT-4Report} version 5. 
Graphs are prepared using Matplotlib~\cite{Hunter2007Matplotlib:Environment}. 
Crystal structures prepared using VESTA 3~\cite{Momma2011VESTA3Data}. 

\textcolor{black}{
\section*{Data availability}
}

\textcolor{black}{
Upon publication, data from the Monte Carlo simulations, and associated Python scripts, will be made available in a data repository managed by the University of Cambridge~\cite{naglecocco_dutton_cambridge_dataset}. 
}

\begin{acknowledgments}

The authors thank James M. A. Steele, Nicola D. Kelly, and Amber Visser at the University of Cambridge, and Annalena R. Genreith-Schriever at RWTH Aachen University, for useful comments.

\end{acknowledgments}

\appendix

\section{Simulated annealing}

In the simulated annealing procedure, temperature is initialised at $T_0=\max{(T_\mathrm{max},10\cdot T_\mathrm{target})}$ where $T_\mathrm{target}$ is the target temperature of the run and $T_\mathrm{max}>>1$. At each iteration $i$ out of a total number of iterations $n$, the temperature $T_i$ is given by:

\begin{equation}\label{simulated_annealing_equation}
    T_{\mathrm{i}} =
    \begin{cases}
        T_0
        \left(\dfrac{T_{\mathrm{target}}}{T_0}\right)^{\tfrac{i}{\,n_{\mathrm{sa}}-1}} 
        & \text{if } i < n_\mathrm{sa}, \\[8pt]
        T_{\mathrm{target}} 
        & \text{if } i \geq n_\mathrm{sa} .
    \end{cases}
\end{equation}

where $n_\mathrm{sa}=0.8n$ rounded down to the nearest integer. 


\bibliography{references}

\onecolumngrid
\clearpage
\newpage
\section*{Supplementary Information} 
\renewcommand{\thesection}{}
\renewcommand{\thesubsection}{S\arabic{subsection}}
\renewcommand{\theequation}{S\arabic{equation}}
\renewcommand{\thefigure}{S\arabic{figure}}
\renewcommand{\thetable}{S\arabic{table}}
\setcounter{figure}{0}
\setcounter{table}{0}
\setcounter{equation}{0}

\listoftables
\listoffigures
\clearpage

\subsection{Other reported orderings from literature}

In this section, some other orbital orderings reported in the literature are presented, to support the discussion throughout the introduction and results of the main text.

\begin{figure}[h]
    \includegraphics[scale=1]{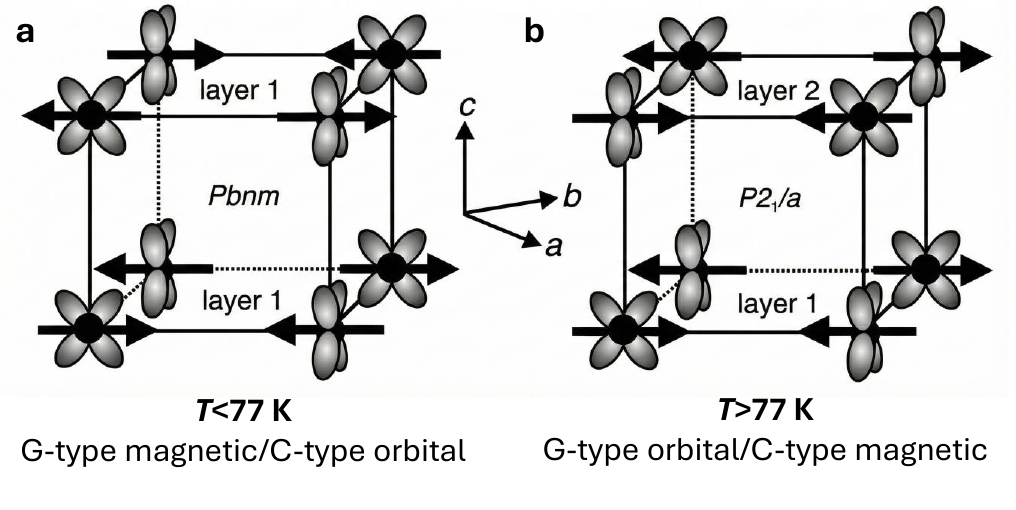}
    \caption[G-type orbital order as reported in literature]{
        \label{SI_G-type_from-lit.pdf}
        The orbital ordering in the perovskite YVO$_3$ at low (left) and high (right) temperatures, representing C- and G-type orbital order respectively, as reported by Blake \textit{et al.}~\cite{Blake2002NeutronYVO3}  
        The C-type orbital order (left) matches that of LaMnO$_3$, corresponding to phase 5 of the phase diagram of Ahmed and Gehring~\cite{Ahmed2006PottsLaMnO3}. 
        The G-type orbital order cannot be recovered from the phase diagram of Ahmed and Gehring. 
    }
\end{figure}

Figure~\ref{SI_G-type_from-lit.pdf} shows the transition between G-type and C-type orbital order reported~\cite{Blake2001TransitionYVO3,Blake2002NeutronYVO3} for YVO$_3$. 
C-type orbital order is studied in this present work; G-type orbital order is not obtained in the phase diagram of Ahmed and Gehring and would likely require an extended model including additional $J$ terms. 
Both orbital orderings shown here are not strictly Jahn--Teller distortions (although $d^2$ vanadates do contain electronic degeneracy).

\begin{figure}[h]
    \includegraphics[width=\linewidth]{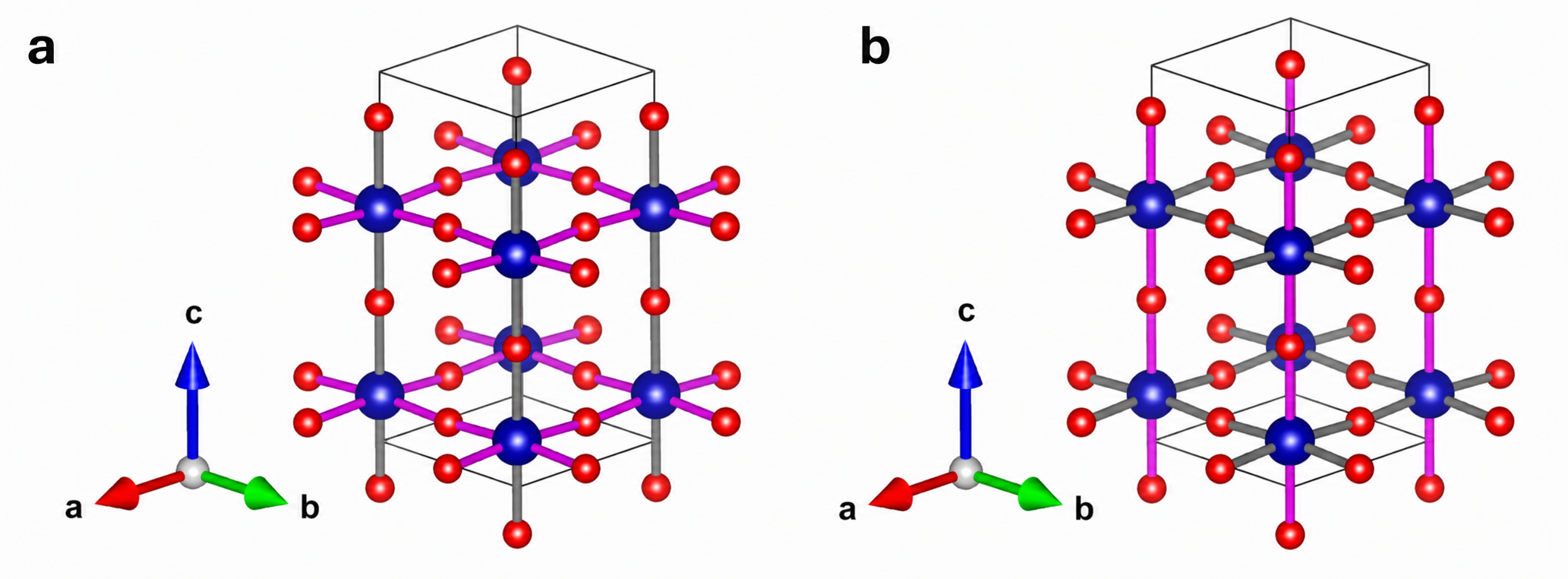}
    \caption[Collinear orbital order in perovskites as reported in literature]{
        \label{Ba0.5La0.5CoO3_order_SI}
        The orbital ordering of the JT distortion in the perovskite La$_{0.5}$Ba$_{0.5}$CoO$_3$, as reported for (a) ordered~\cite{Nakajima2005NewEffect} and (b) disordered~\cite{Fauth2002IntermediateDistortions} A-site cations. 
        In both schematics, the crystal structure is viewed along the [111] direction. 
        Here, A-site cations are excluded. 
        Blue atoms represent Co cations and red atoms reperesent O anions. 
        In terms of bonds, grey bonds are short and pink bonds are long. 
    }
\end{figure}

Figure~\ref{Ba0.5La0.5CoO3_order_SI} shows the two reported orbital orderings of La$_{0.5}$Ba$_{0.5}$CoO$_6$ perovskite~\cite{Fauth2002IntermediateDistortions,Nakajima2005NewEffect}, which differ depending on whether the A site is ordered or disordered. 
Both are collinear ordering, but JT distortions are compressive and elongational respectively. It should be noted that the compressive structure~\cite{Nakajima2005NewEffect} exhibits some $Q_2$ component which is not a degree of freedom in our simulations.

\begin{figure}[h]
    \includegraphics[scale=1]{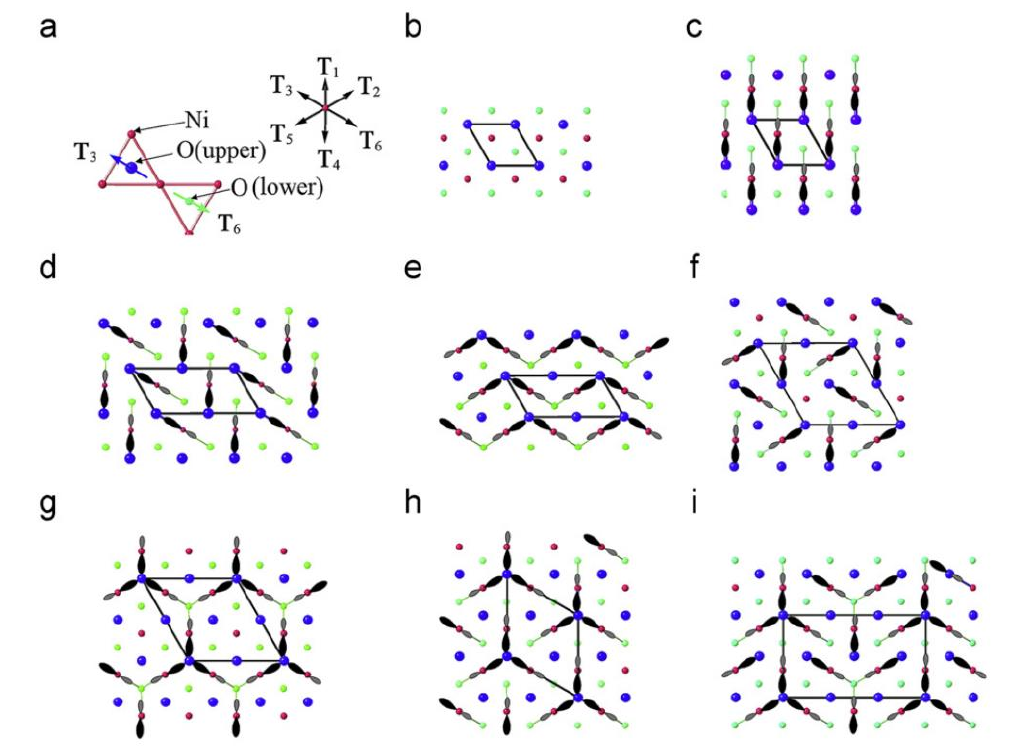}
    \caption[Various proposed layered nickelate orbital order as reported in literature]{
        \label{SI_LNO-ordering-from-lit.pdf}
        Various proposed models for layered nickelate orbital ordering. 
        Models viewed along the $c$ axis. The red balls denote Ni ions, the blue and green balls are oxygen ions in the upper plane and in the lower plane, respectively. 
        Only the shifted oxygen ions are linked with the Ni ions in the sketches. 
        The unit cell of each model is outlined with black lines. 
        (a) Definitions of the six shift directions in the flat oxygen plane: $T_{1}=a+2b$; $T_{2}=2a+b$; $T_{3}=-a+b$; $T_{4}=-a-2b$; $T_{5}=-2a-b$; $T_{6}=a-b$. Vectors $a$, $b$ and $c$ are the basis vectors of the $R\bar{3}m$ hexagonal lattice. 
        (b) $R\bar{3}m$. 
        (c) $C2/m$. (d) Zigzag. (e) Dimer. (f) Windmill. 
        (g) Honeycomb. (h) Trimer. (i) Alternating trimer. 
        Reprinted from Journal of Solid State Chemistry, Vol. 184, Issue 7, Chen, Zhenlian; Zou, Huamin; Zhu, Xiaopeng; Zou, Jie; Cao, Jiefeng. \textit{First-principle investigation of Jahn–Teller distortion and topological analysis of chemical bonds in LiNiO$_2$}, pp. 1784–1790, Copyright (2011), with permission from Elsevier.~\cite{Chen2011First-principleLiNiO2}.
    }
\end{figure}

Figure~\ref{SI_LNO-ordering-from-lit.pdf} shows various proposed orbital orderings for LiNiO$_2$, including the collinear structure which we find as being the ideal structure within domains, and the trimers and dimers which we observe at the interfaces of orbitally-ordered domains. 

\begin{figure}[h]
    \includegraphics[scale=0.4]{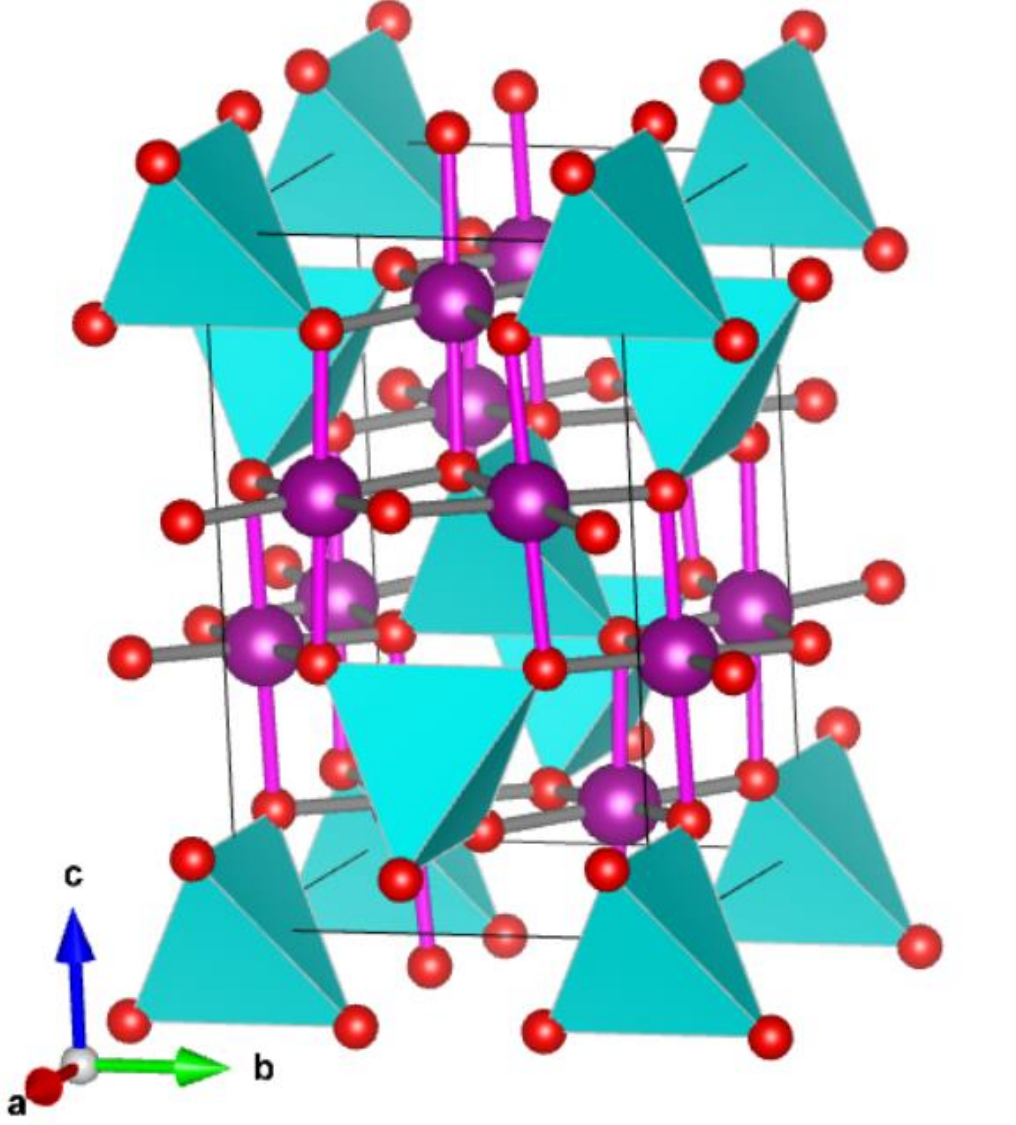}
    \caption[Collinear orbital order in spinels from literature]{
        \label{SI_spinel_ZnMn2O4_literature.pdf}
        Proposed collinear orbital ordering in spinels with ZnMn$_2$O$_4$~\cite{Patra2019StructuralZnMn2O4} taken as an example. Pink bonds indicate the axes of elongation of octahedra. 
    }
\end{figure}

Orbital ordering in spinels is almost invariably found to be collinear. Figure~\ref{SI_spinel_ZnMn2O4_literature.pdf} shows a depiction of the octahedral elongations in this ordering. 

\clearpage
\subsection{Reproducing the anisotropic Potts model perovskite phase diagram with static $\rho$}\label{J1_J2_reproduction_section}

\begin{figure}[h]
    \includegraphics[width=\linewidth]{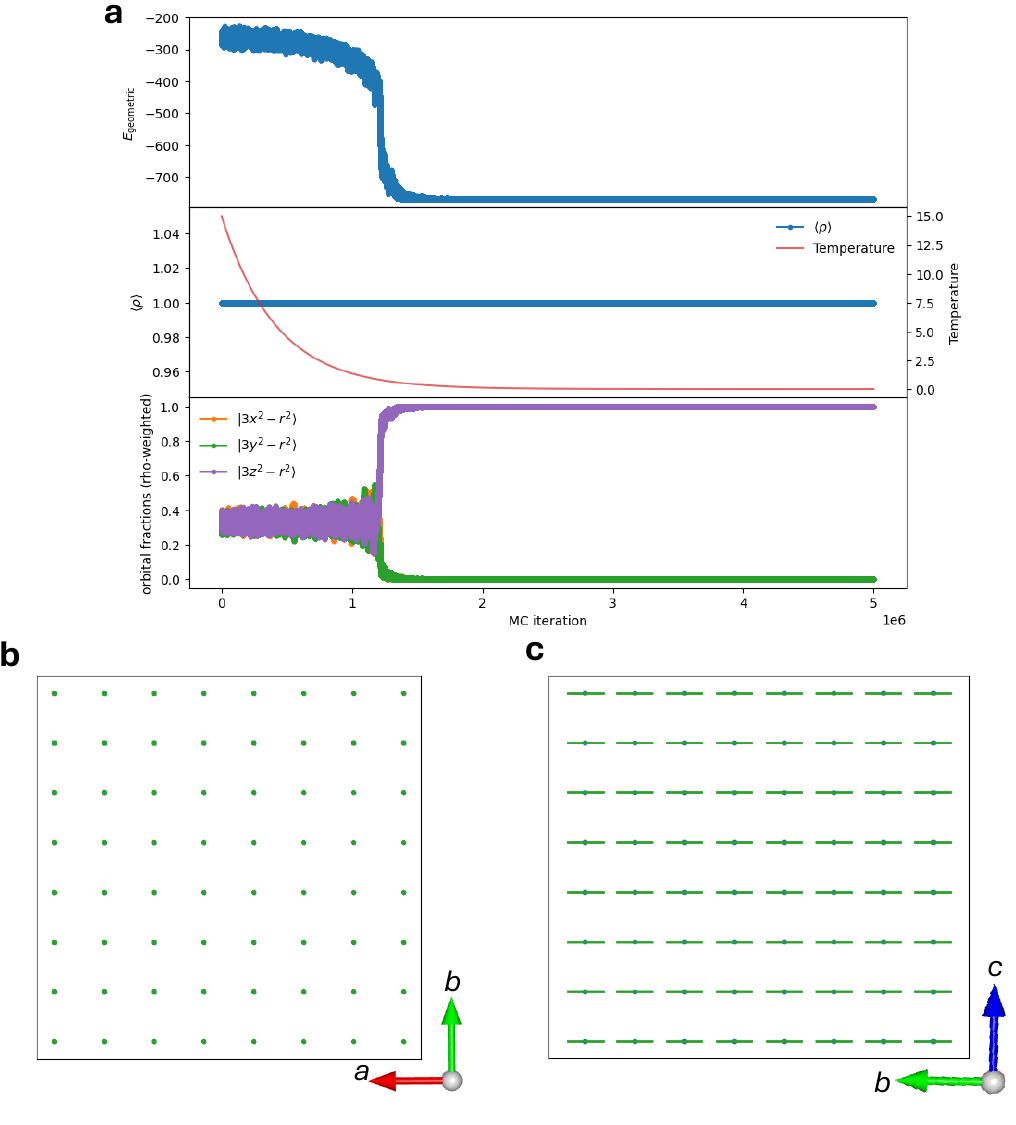}
    \caption[Summary of Monte Carlo perovskite run with static $\rho$ to reproduce phase 1 of ($J_1,J_2$) phase diagram]{
        \label{figure_static-rho_phase1}
        The results of a Monte Carlo \textcolor{black}{simulation} with static $\rho=1$ ($P_\mathrm{switch}=0$), reproducing Phase 1 of the Ahmed and Gehring~\cite{Ahmed2005TheModel} phase diagram for the anisotropic Potts model in a perovskite. 
        (a) Energy with iteration (top), mean $\rho$ and temperature against iteration (middle), fractional occupations of $e_g$ orbitals in each direction (bottom). 
        \textcolor{black}{$E_\mathrm{single-ion}$ is not plotted as it does not deviate from zero in static mode.}
        (b) Example configuration in the $ab$-plane in a randomly-selected cross-section. 
        (c) Example configuration in the $bc$-plane in a randomly-selected cross-section. 
        Energy parameters $\alpha=\beta=0$ from equation~\ref{hamiltonian_equation}. 
        Simulated annealing was used before settling on a final $T=0.001$. 
        This run lasted for $5\times10^6$ iterations.
    }
\end{figure}

\begin{figure}[h]
    \includegraphics[width=\linewidth]{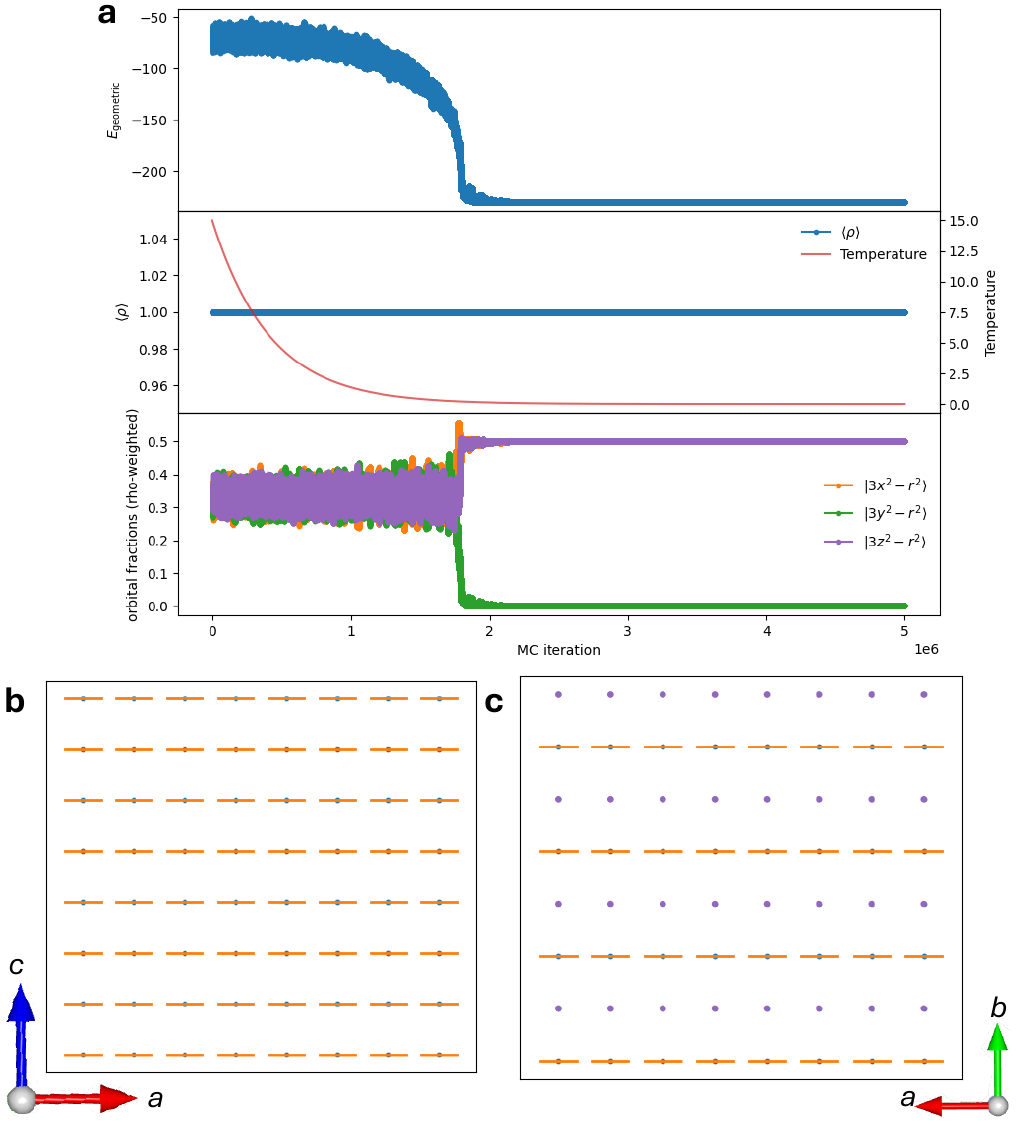}
    \caption[Summary of Monte Carlo perovskite run with static $\rho$ to reproduce phase 2 of ($J_1,J_2$) phase diagram]{
        \label{figure_static-rho_phase2}
        The results of a Monte Carlo \textcolor{black}{simulation} with static $\rho=1$ ($P_\mathrm{switch}=0$), reproducing Phase 2 of the Ahmed and Gehring~\cite{Ahmed2005TheModel} phase diagram for the anisotropic Potts model in a perovskite. 
        (a) Energy with iteration (top), mean $\rho$ and temperature against iteration (middle), fractional occupations of $e_g$ orbitals in each direction (bottom). 
        \textcolor{black}{$E_\mathrm{single-ion}$ is not plotted as it does not deviate from zero in static mode.}
        (b) Example configuration in the $ac$-plane in a randomly-selected cross-section. 
        (c) Example configuration in the $ab$-plane in a randomly-selected cross-section. 
        Energy parameters $\alpha=\beta=0$ from equation~\ref{hamiltonian_equation}. 
        Simulated annealing was used before settling on a final $T=0.001$. 
        This run lasted for $5\times10^6$ iterations.
    }
\end{figure}

\begin{figure}[h]
    \includegraphics[width=\linewidth]{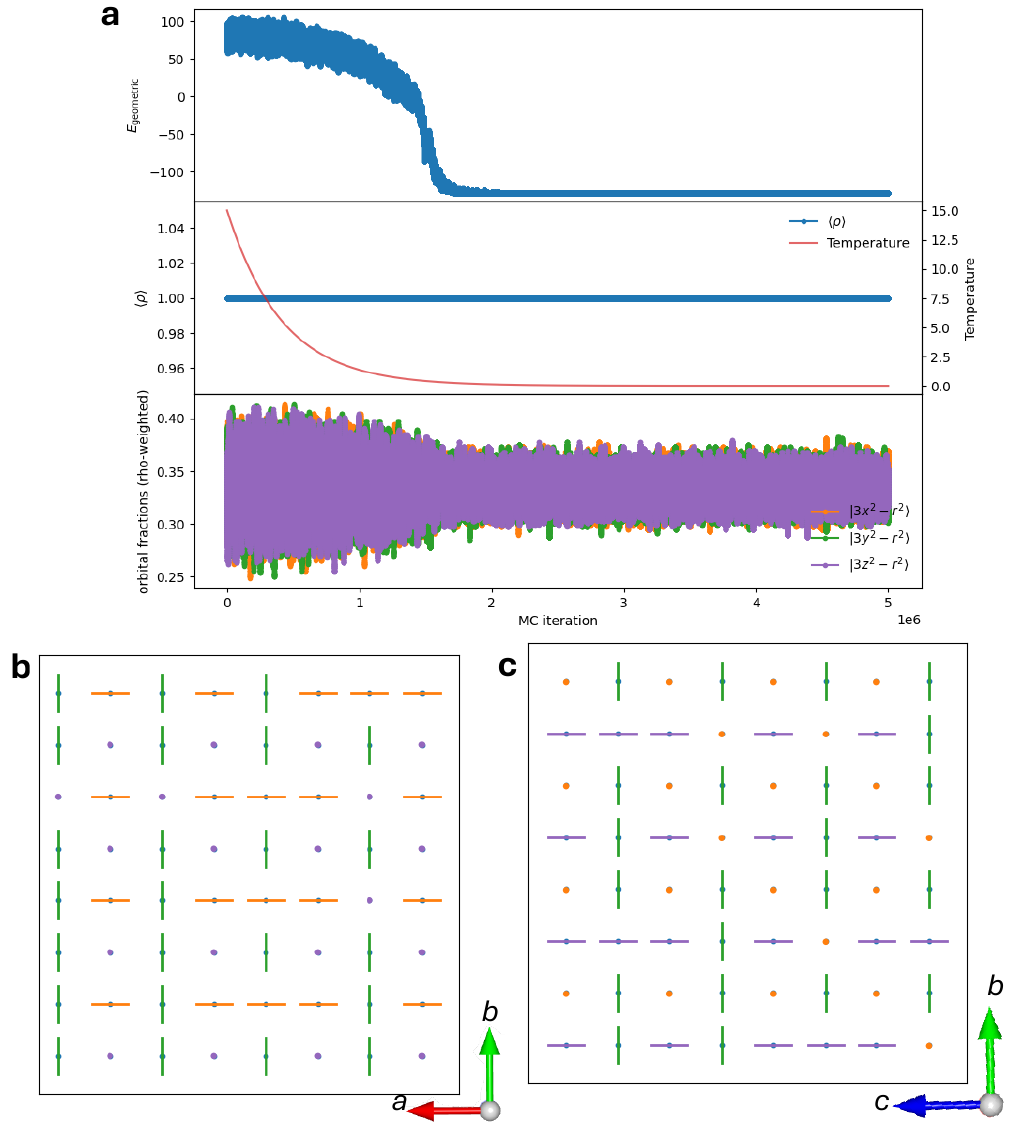}
    \caption[Summary of Monte Carlo perovskite run with static $\rho$ to reproduce phase 3 of ($J_1,J_2$) phase diagram]{
        \label{figure_static-rho_phase3}
        The results of a Monte Carlo \textcolor{black}{simulation} with static $\rho=1$ ($P_\mathrm{switch}=0$), reproducing Phase 3 of the Ahmed and Gehring~\cite{Ahmed2005TheModel} phase diagram for the anisotropic Potts model in a perovskite. 
        (a) Energy with iteration (top), mean $\rho$ and temperature against iteration (middle), fractional occupations of $e_g$ orbitals in each direction (bottom). 
        \textcolor{black}{$E_\mathrm{single-ion}$ is not plotted as it does not deviate from zero in static mode.}
        (b) Example configuration in the $ab$-plane in a randomly-selected cross-section. 
        (c) Example configuration in the $bc$-plane in a randomly-selected cross-section. 
        Energy parameters $\alpha=\beta=0$ from equation~\ref{E_single-ion_term}. 
        Simulated annealing was used before settling on a final $T=0.001$. 
        This run lasted for $5\times10^6$ iterations. 
        While there are clearly defects in the structure relative to the phase 3 ``cage" ground state, the proposed ``cage"-ordering is broadly maintained. 
    }
\end{figure}

\begin{figure}[h]
    \includegraphics[width=\linewidth]{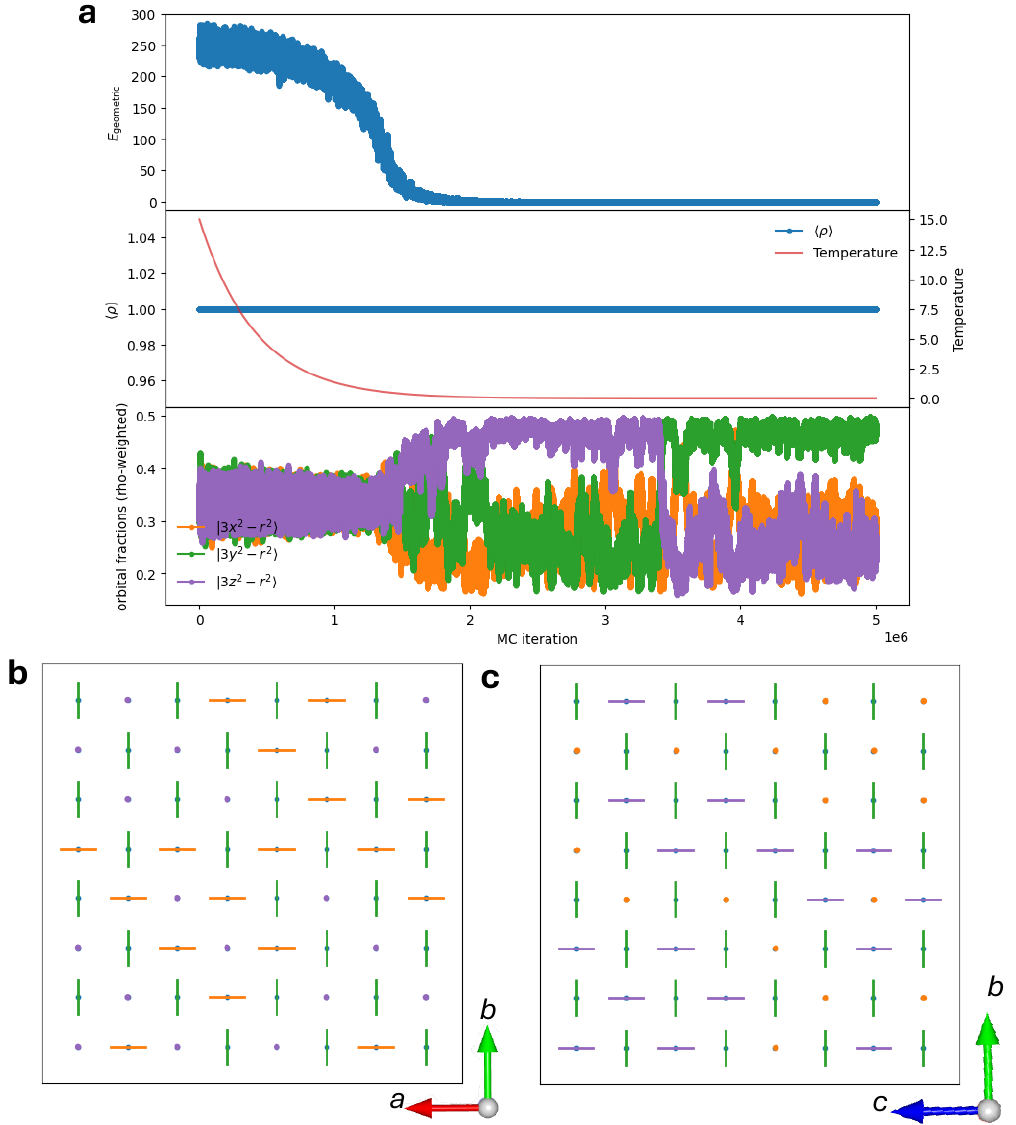}
    \caption[Summary of Monte Carlo perovskite run with static $\rho$ to reproduce phase 4 of ($J_1,J_2$) phase diagram]{
        \label{figure_static-rho_phase4}
        The results of a Monte Carlo \textcolor{black}{simulation} with static $\rho=1$ ($P_\mathrm{switch}=0$), reproducing the antiferromagnetic Phase 4 of the Ahmed and Gehring~\cite{Ahmed2005TheModel} phase diagram for the anisotropic Potts model in a perovskite. 
        (a) Energy with iteration (top), mean $\rho$ and temperature against iteration (middle), fractional occupations of $e_g$ orbitals in each direction (bottom). 
        \textcolor{black}{$E_\mathrm{single-ion}$ is not plotted as it does not deviate from zero in static mode.}
        (b) Example configuration in the $ab$-plane in a randomly-selected cross-section. 
        (c) Example configuration in the $bc$-plane in a randomly-selected cross-section. 
        Energy parameters $\alpha=\beta=0$ from equation~\ref{E_single-ion_term}. 
        Simulated annealing was used before settling on a final $T=0.001$. 
        This run lasted for $5\times10^6$ iterations. 
        The switching behaviour we see even at lowest temperature suggests instability associated with having antiferromagnetic $J_1=J_2$, although the antiferromagnetic behaviour reported by Ahmed and Gehring is essentially reproduced.
    }
\end{figure}

\begin{figure}[h]
    \includegraphics[width=\linewidth]{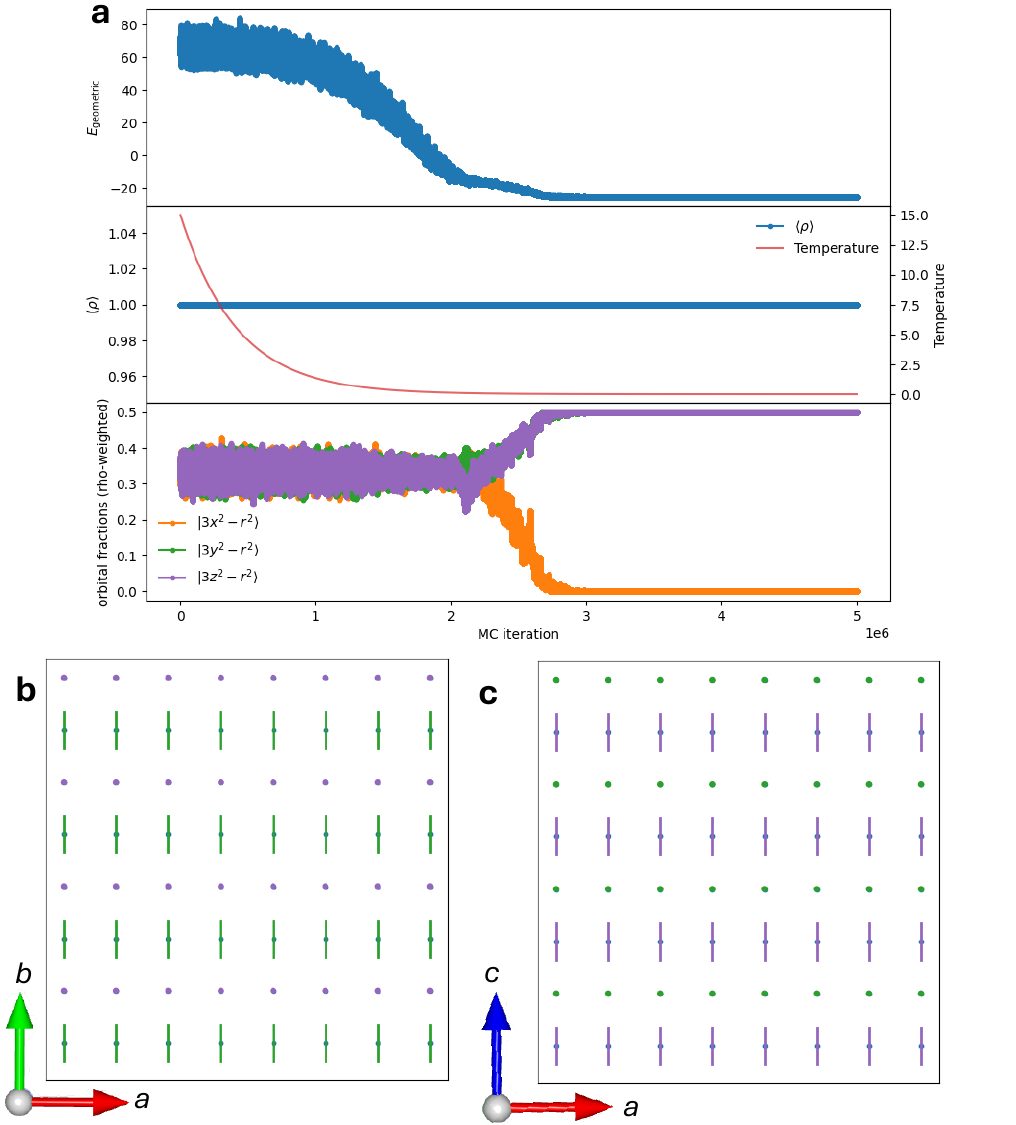}
    \caption[Summary of Monte Carlo perovskite run with static $\rho$ to reproduce phase 5 of ($J_1,J_2$) phase diagram]{
        \label{figure_static-rho_phase5}
        The results of a Monte Carlo \textcolor{black}{simulation} with static $\rho=1$ ($P_\mathrm{switch}=0$), reproducing Phase 5 of the Ahmed and Gehring~\cite{Ahmed2005TheModel} phase diagram for the anisotropic Potts model in a perovskite. 
        (a) Energy with iteration (top), mean $\rho$ and temperature against iteration (middle), fractional occupations of $e_g$ orbitals in each direction (bottom). 
        \textcolor{black}{$E_\mathrm{single-ion}$ is not plotted as it does not deviate from zero in static mode.}
        (b) Example configuration in the $ab$-plane in a randomly-selected cross-section. 
        (c) Example configuration in the $ac$-plane in a randomly-selected cross-section. 
        Energy parameters $\alpha=\beta=0$ from equation~\ref{E_single-ion_term}. 
        Simulated annealing was used before settling on a final $T=0.001$. 
        This run lasted for $5\times10^6$ iterations. 
        We note the presence of a defect here. As seen in the $ab$-plane, there is a single collinear plane which has the effect that $\sim$12.5\% of sites have an elongation in $|3y^2-r^2>$, even though the energy is mimimised analytically \textit{via} the ordering in the other 7 (of 8) planes. 
        With slower annealing we could likely remove this defect, but it is instructive to keep it in place to show the tendency of Monte Carlo simulations towards local minima at low temperatures, even with simulated annealing. 
    }
\end{figure}

\begin{figure}[h]
    \includegraphics[width=\linewidth]{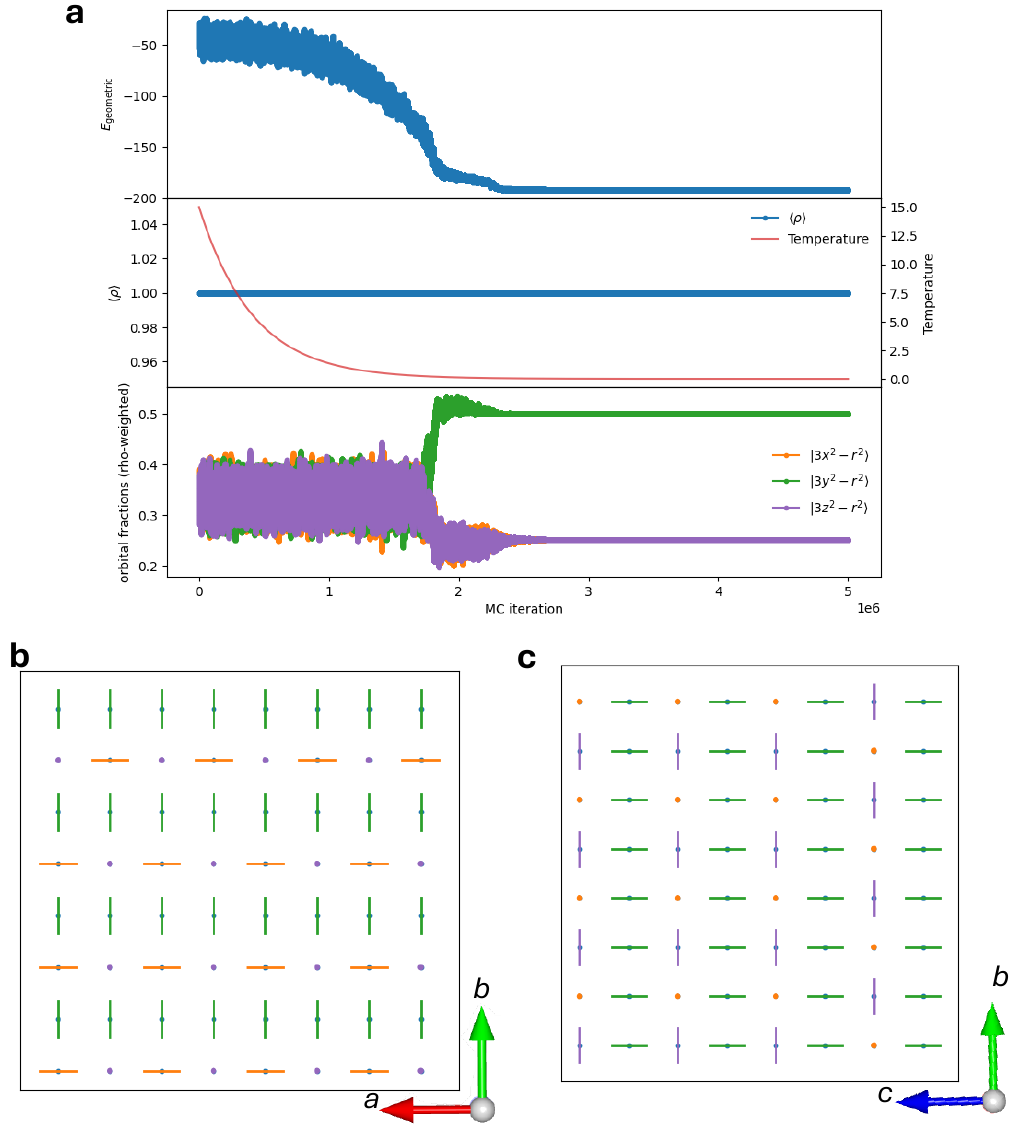}
    \caption[Summary of Monte Carlo perovskite run with static $\rho$ to reproduce phase 6 of ($J_1,J_2$) phase diagram]{
        \label{figure_static-rho_phase6}
        The results of a Monte Carlo \textcolor{black}{simulation} with static $\rho=1$ ($P_\mathrm{switch}=0$), reproducing Phase 6 of the Ahmed and Gehring~\cite{Ahmed2005TheModel} phase diagram for the anisotropic Potts model in a perovskite. 
        (a) Energy with iteration (top), mean $\rho$ and temperature against iteration (middle), fractional occupations of $e_g$ orbitals in each direction (bottom). 
        \textcolor{black}{$E_\mathrm{single-ion}$ is not plotted as it does not deviate from zero in static mode.} 
        (b) Example configuration in the $ab$-plane in a randomly-selected cross-section. 
        (c) Example configuration in the $bc$-plane in a randomly-selected cross-section. 
        Energy parameters $\alpha=\beta=0$ from equation~\ref{E_single-ion_term}. 
        Simulated annealing was used before settling on a final $T=0.001$. 
        This run lasted for $5\times10^6$ iterations. 
    }
\end{figure}

To validate our code, we tested our code using the specific conditions matching that of Ahmed and Gehring's Monte Carlo simulations~\cite{Ahmed2005TheModel,Ahmed2006PottsLaMnO3}. In this case, $\rho$ is fixed to 1 for all JT-active sites, and $P_\mathrm{switch}=0$ to ensure there is no variation in the JT activity of each site. 
This is a specific case of the Hamiltonian in Equation~\ref{hamiltonian_equation} where $\alpha=\beta=0$. 
Table~\ref{Table_phases_ahmed_gehring_static-rho} lists the phases in the anisotropic Potts phase diagram which were identified by Ahmed and Gehring~\cite{Ahmed2005TheModel}, along with the conditions by which we reproduced this phase. 
Figures~\ref{figure_static-rho_phase1} to \ref{figure_static-rho_phase6} shows the results of Monte Carlo runs under these conditions to reproduce the six phases. 
We find that overall the phases are reproduced, but often we had to use several runs with varying $T_\mathrm{max}$ to reproduce the expected ordering. 

\begin{table}[h!]
\centering
\begin{tabular}{|c|c|c|c|c|c|}
\hline
\textbf{Phase} & \textbf{Constraints in $( J_1, J_2)$} & Values tested & \multicolumn{2}{c}{Static $\rho$ test} & Dynamic $\rho$ test \\
& & & $T_\mathrm{max}$ & Figure & \\
\hline
1 & $ J_2 < 0,\;\;  J_1 < - J_2$ & $J_2=J_1=-1$ & 100 & Figure~\ref{figure_static-rho_phase1} & Figure~\ref{figure_dynamic-rho_phase1}\\
\hline
2 & $ J_2 > 0,\;\; J_1 < - 2 J_2$ & $J_1=-1$, $J_2=0.1$ & 15 & Figure~\ref{figure_static-rho_phase2} & -\\
\hline
3 & $ J_2 > 0,\;\;  0 > J_1 > -2 J_2$ & $J_1=-1$, $J_2=1$ & 15 & Figure~\ref{figure_static-rho_phase3} & -\\
\hline
4 & $ J_2 > 0,\;\;  J_1 > 0$ & $J_1=J_2=1$ & 15 & Figure~\ref{figure_static-rho_phase4} & -\\
\hline
5 & $ J_2 < 0,\;\; J_1 > - 2 J_2$ & $J_1=1$, $J_2=-0.1$ & 15 & Figure~\ref{figure_static-rho_phase5} & Figures~\ref{perovskite_C-type_lowT_simulation.pdf} and \ref{figure_dynamic-rho_phase5}\\
\hline
6 & $ J_2 < 0,\;\; -2 J_2 > J_1 > - J_2$ & $J_1=1$, $J_2=-0.75$ & 100 & Figure~\ref{figure_static-rho_phase6} & - \\
\hline
\end{tabular}
\caption[Unique phases in the ($J_1$,$J_2$) phase diagram for perovskites]{
    \label{Table_phases_ahmed_gehring_static-rho}
    Phases of the anisotropic Potts model reported by Ahmed and Gehring~\cite{Ahmed2005TheModel,Ahmed2006PottsLaMnO3}. 
    This is a specific case of the Hamiltonian in Equation~\ref{hamiltonian_equation} where $\alpha=\beta=0$, and in the simulations $P_\mathrm{switch}=0$. 
    Simulated annealing was used with variable $T_\mathrm{max}$, as described in Equation~\ref{simulated_annealing_equation}, before settling on a final $T=0.001$. 
    Note that we use opposite sign conventions for $J_1$ and $J_2$ compared with Ahmed and Gehring. 
    Each run lasted for $5\times10^6$ iterations. 
}
\end{table}

\begin{figure}[h]
    \includegraphics[scale=1]{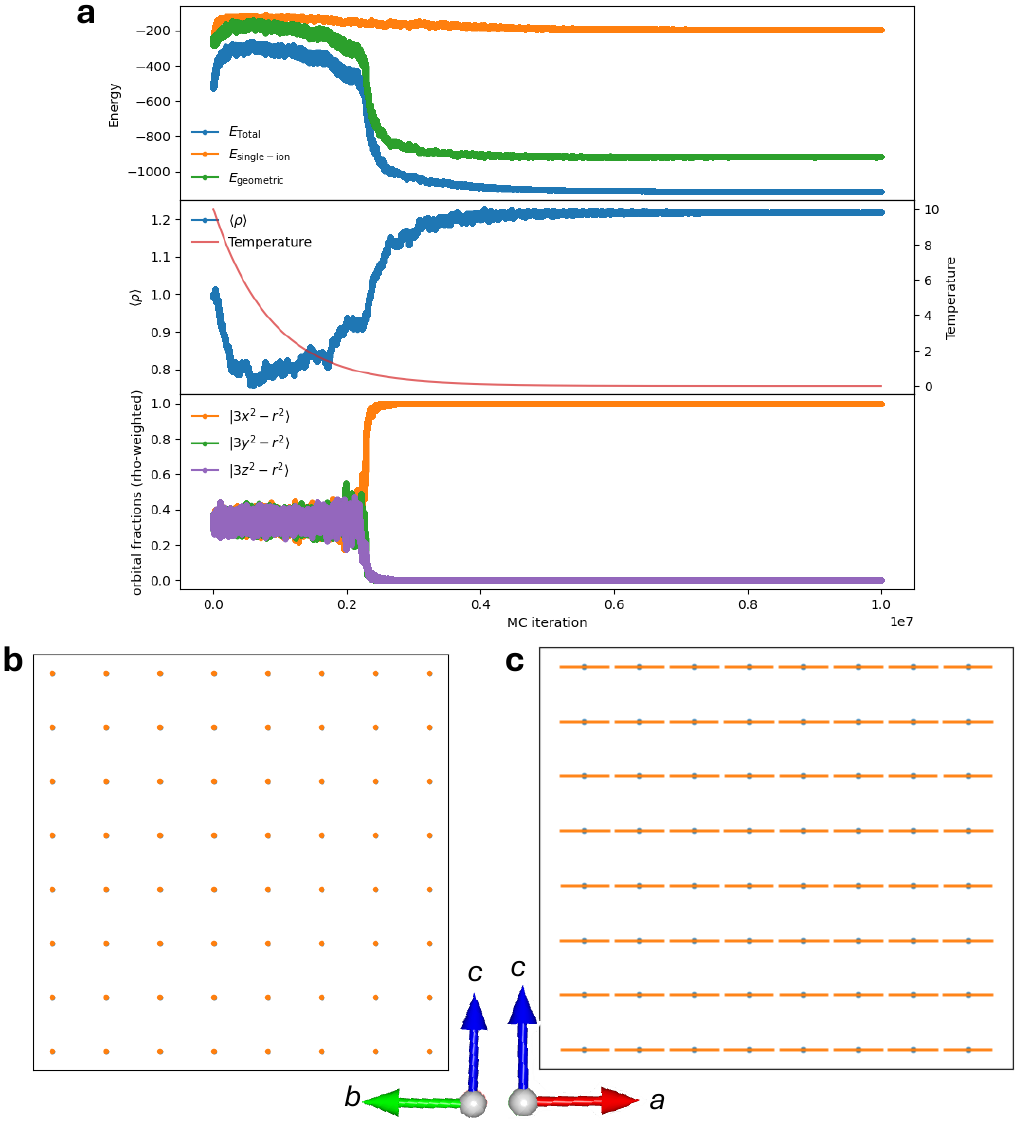}
    \caption[Summary of Monte Carlo perovskite run with dynamic $\rho$ to reproduce phase 1 of ($J_1,J_2$) phase diagram]{
        \label{figure_dynamic-rho_phase1}
        The results of a Monte Carlo \textcolor{black}{simulation} with dynamic $\rho$ ($P_\mathrm{switch}=1/2$), reproducing Phase 1 of the Ahmed and Gehring~\cite{Ahmed2005TheModel} phase diagram for the anisotropic Potts model in a perovskite. 
        (a) Energy with iteration (top), mean $\rho$ and temperature against iteration (middle), fractional occupations of $e_g$ orbitals in each direction (bottom).
        (b) Example configuration in the $bc$-plane in a randomly-selected cross-section. 
        (c) Example configuration in the $ac$-plane in a randomly-selected cross-section. 
        Energy parameters $\alpha=-1$,$\beta=-1$  from equation~\ref{E_single-ion_term}. 
        Simulated annealing was used before settling on a final $T=0.001$. 
        This run lasted for $10^7$ iterations.
    }
\end{figure}

\begin{figure}[h]
    \includegraphics[scale=1]{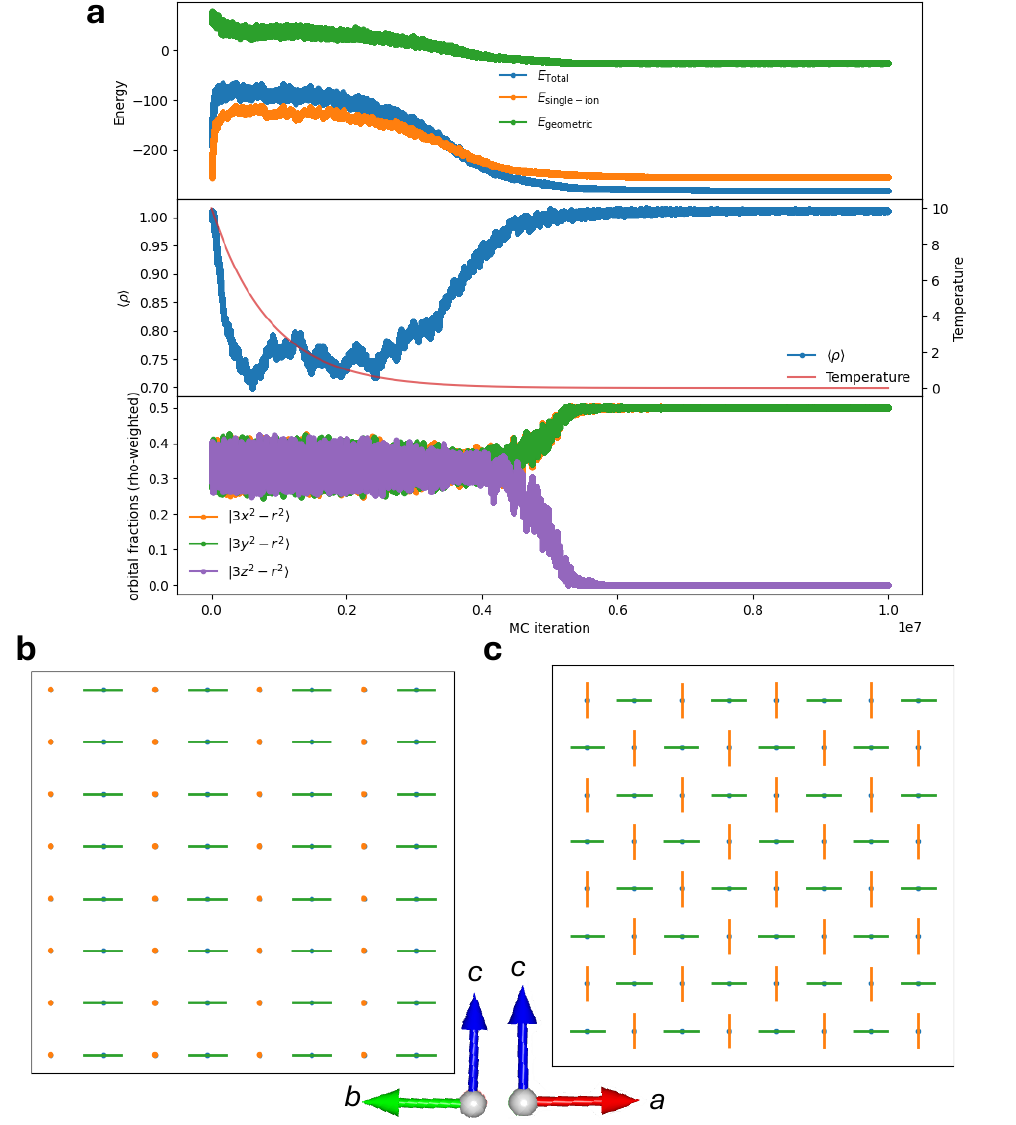}
    \caption[Summary of Monte Carlo perovskite run with dynamic $\rho$ to reproduce phase 5 of ($J_1,J_2$) phase diagram]{
        \label{figure_dynamic-rho_phase5}
        The results of a Monte Carlo \textcolor{black}{simulation} with dynamic $\rho$ ($P_\mathrm{switch}=1/2$), reproducing Phase 5 of the Ahmed and Gehring~\cite{Ahmed2005TheModel} phase diagram for the anisotropic Potts model in a perovskite. 
        (a) Energy with iteration (top), mean $\rho$ and temperature against iteration (middle), fractional occupations of $e_g$ orbitals in each direction (bottom).
        (b) Example configuration in the $bc$-plane in a randomly-selected cross-section. 
        (c) Example configuration in the $ac$-plane in a randomly-selected cross-section. 
        Energy parameters $\alpha=-1$,$\beta=1/2$  from equation~\ref{E_single-ion_term}. 
        Simulated annealing was used before settling on a final $T=0.001$. 
        This run lasted for $10^7$ iterations.
    }
\end{figure}

Beyond this, we then tested a dynamic $\rho$ model, in which $\alpha \ne 0$ and $\beta \ne 0$, and $P_\mathrm{switch}=0.5$, to ensure our extended Hamiltonian in Equation~\ref{hamiltonian_equation} can also reproduce the phase diagram. 
We test this on phases 1 [Figure~\ref{figure_dynamic-rho_phase1}] and 5 [Figures~\ref{perovskite_C-type_lowT_simulation.pdf} and \ref{figure_dynamic-rho_phase5}].


\clearpage
\subsection{Variable-size testing for perovskite cells}

\begin{figure}[p]
    \includegraphics[scale=1.0]{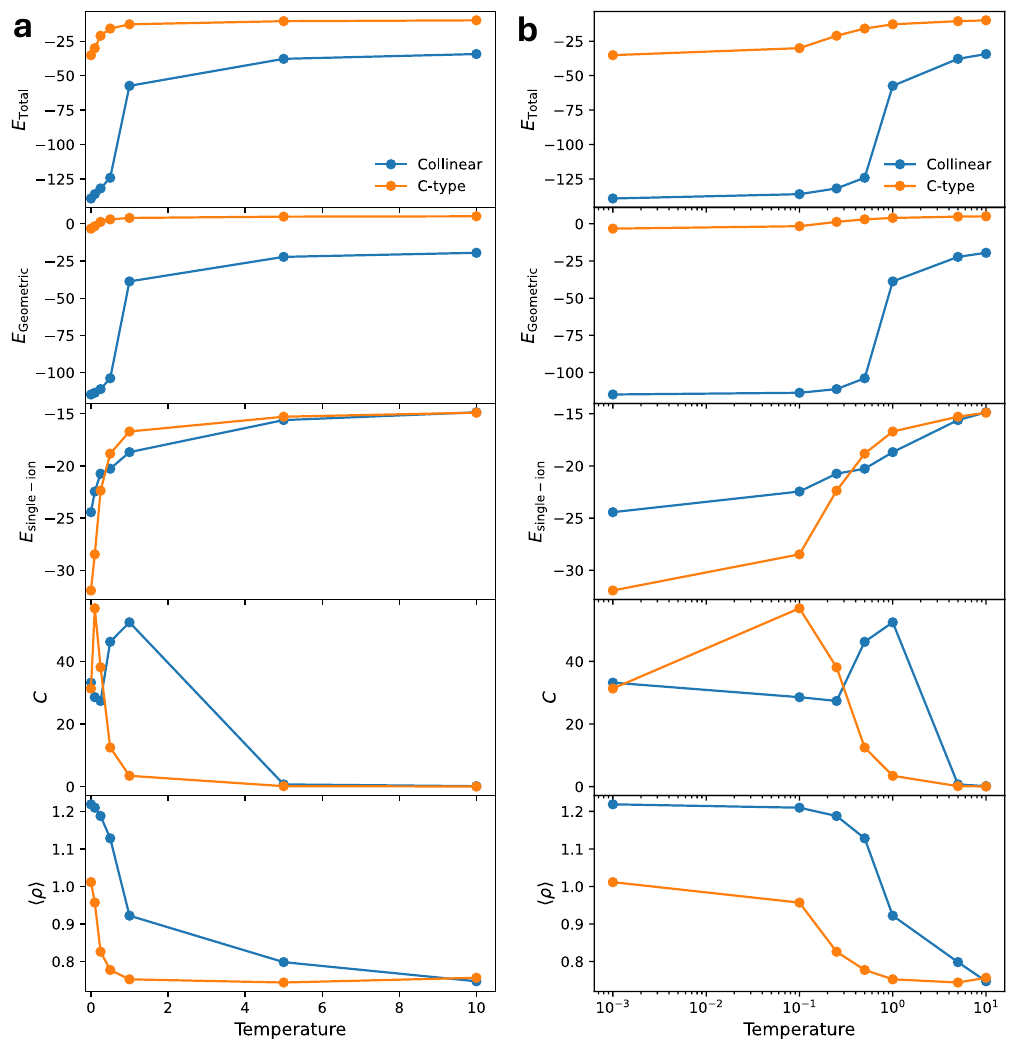}
    \caption[Temperature-dependence of energy, $\langle\rho\rangle$, and $C$ for a $4\times4\times4$ perovskite supercell]{
        \label{SI_size-with-E_perov_4x4x4.pdf}
        Energy terms, $\langle\rho\rangle$, and $C$ with temperature, averaged over the final 10\% of iterations in Monte Carlo simulations, as a function of temperature for the collinear and C-type orbital orderings in a $4\times4\times4$ perovskite lattice.
        Each run lasted for $10^7$ iterations. 
        Data are plotted with temperature on a (a) non-logarithmic and (b) logarithmic scale.
    }
\end{figure}

\begin{figure}[p]
    \includegraphics[scale=1.0]{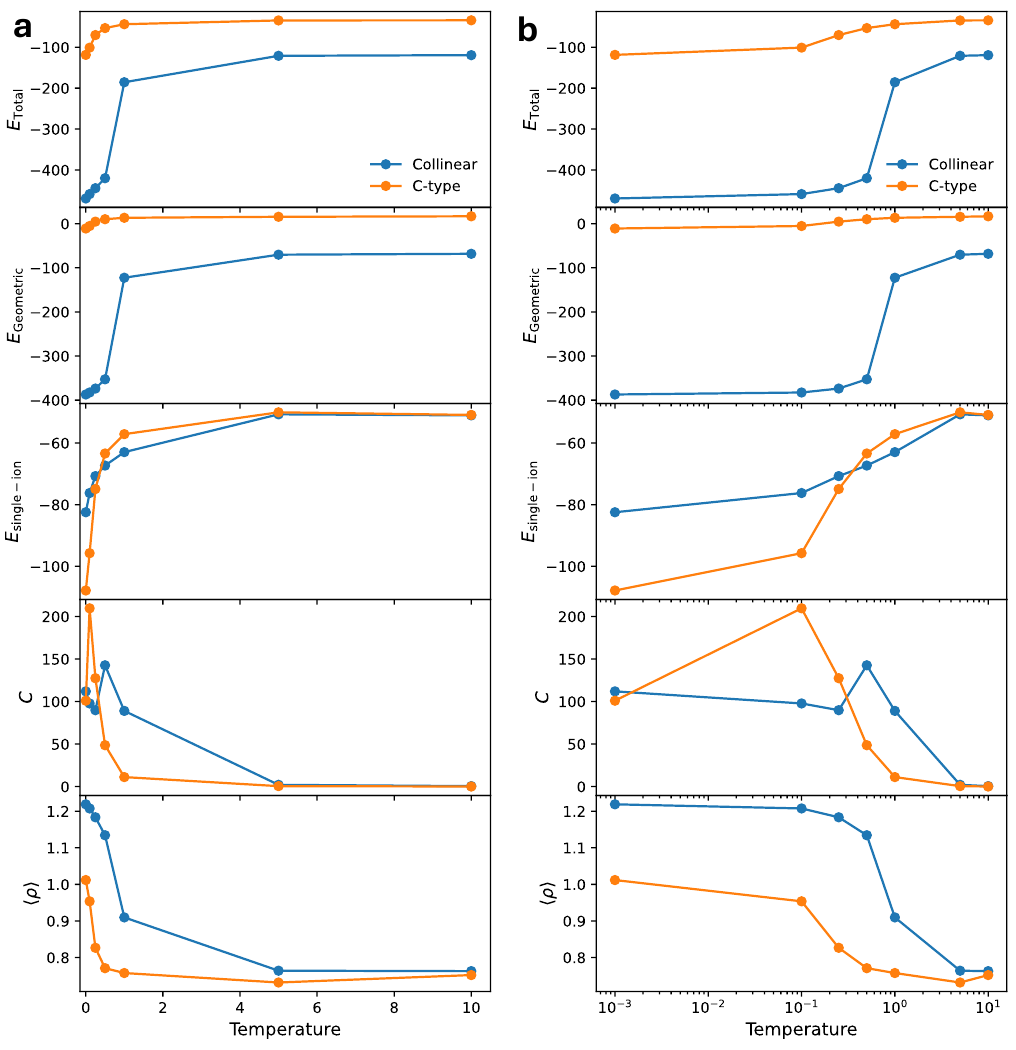}
    \caption[Temperature-dependence of energy, $\langle\rho\rangle$, and $C$ for a $6\times6\times6$ perovskite supercell]{
        \label{SI_size-with-E_perov_6x6x6.pdf}
        Energy terms, $\langle\rho\rangle$, and $C$ with temperature, averaged over the final 10\% of iterations in Monte Carlo simulations, as a function of temperature for the collinear and C-type orbital orderings in a $6\times6\times6$ perovskite lattice.
        Each run lasted for $10^7$ iterations. 
        Data are plotted with temperature on a (a) non-logarithmic and (b) logarithmic scale.
    }
\end{figure}

To test the impact of supercell size on the results (and ensure the resilience of our findings using the $8\times8\times8$ supercell with the perovskite structure) we also performed measurements using supercells with edge size 4 and 6. 
Figures~\ref{SI_size-with-E_perov_4x4x4.pdf} and \ref{SI_size-with-E_perov_6x6x6.pdf} show the variable-temperature energy, heat capacity, and $\langle\rho\rangle$ for $4\times4\times4$ and $6\times6\times6$ supercells, respectively. 
We did not observe qualitative variation in results with cell size. 

\clearpage
\subsection{Histograms of $\rho$ in final configuration}

In Figure~\ref{nickelate-histograms.pdf} in the main text, we presented histograms of $\rho$ in the final configuration for a $8\times8\times8$ collinear perovskite ordered supercell and a $10\times30$ nickelate layer with $K_\mathrm{oxy}\rightarrow\infty$. 
In this section, we present such histograms for both perovskite orderings studied with temperature, along with all three values of $K_\mathrm{oxy}$ used in this work. 

\begin{figure}[h]
    \includegraphics[scale=1.0]{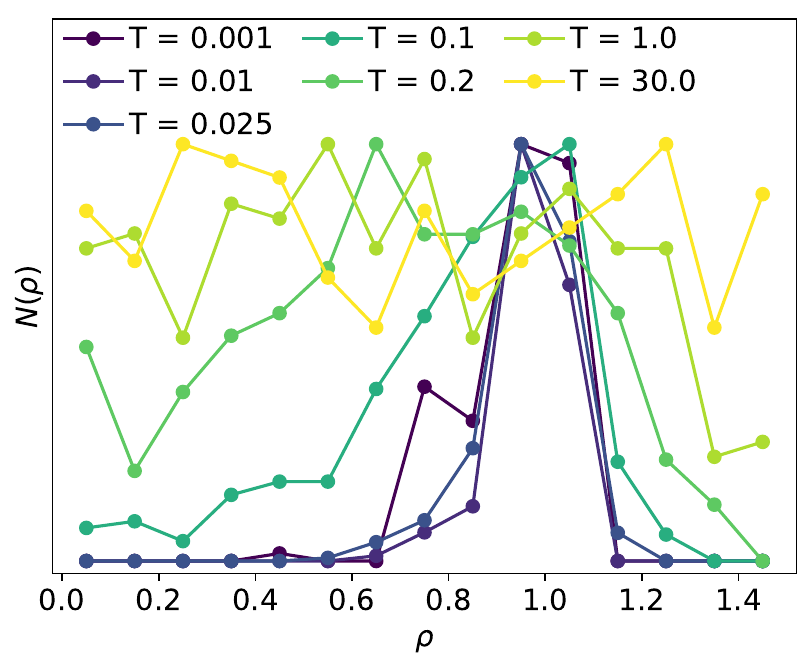}
    \caption[Final histogram of $\rho$ distribution for layered nickelate with $K_\mathrm{oxy}=1$]{
        \label{nickelate_Koxy1_rho_histogram.pdf}
        Final histogram, after $10^7$ iterations, of $\rho$ distribution for layered nickelate with $K_\mathrm{oxy}=1$, at a selection of temperatures. 
        The cell here was a single $10\times30$ nickelate layer.
    }
\end{figure}

\begin{figure}[h]
    \includegraphics[scale=1.0]{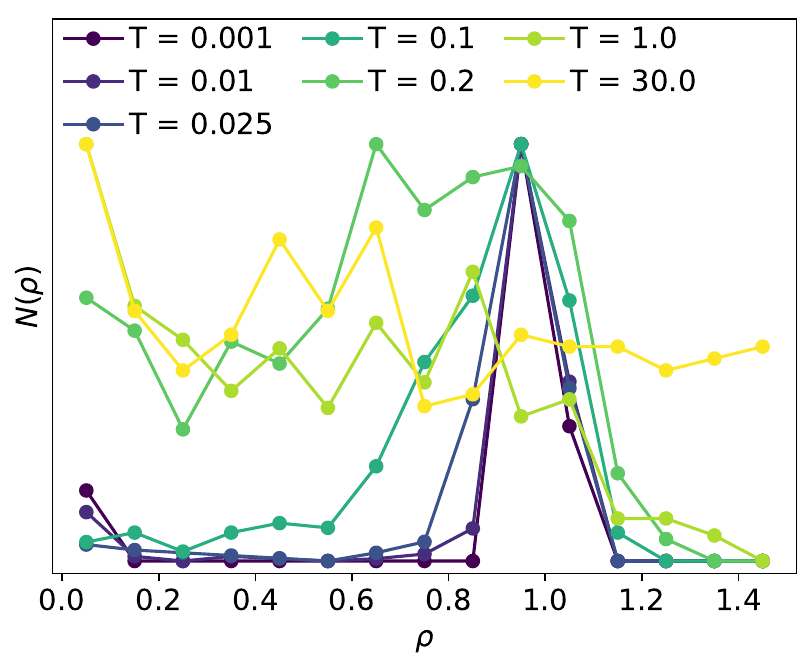}
    \caption[Final histogram of $\rho$ distribution for layered nickelate with $K_\mathrm{oxy}=1$]{
        \label{nickelate_Koxy10_rho_histogram.pdf}
        Final histogram, after $10^7$ iterations, of $\rho$ distribution for layered nickelate with $K_\mathrm{oxy}=10$, at a selection of temperatures. 
        The cell here was a single $10\times30$ nickelate layer.
    }
\end{figure}

\begin{figure}[h]
    \includegraphics[scale=1.0]{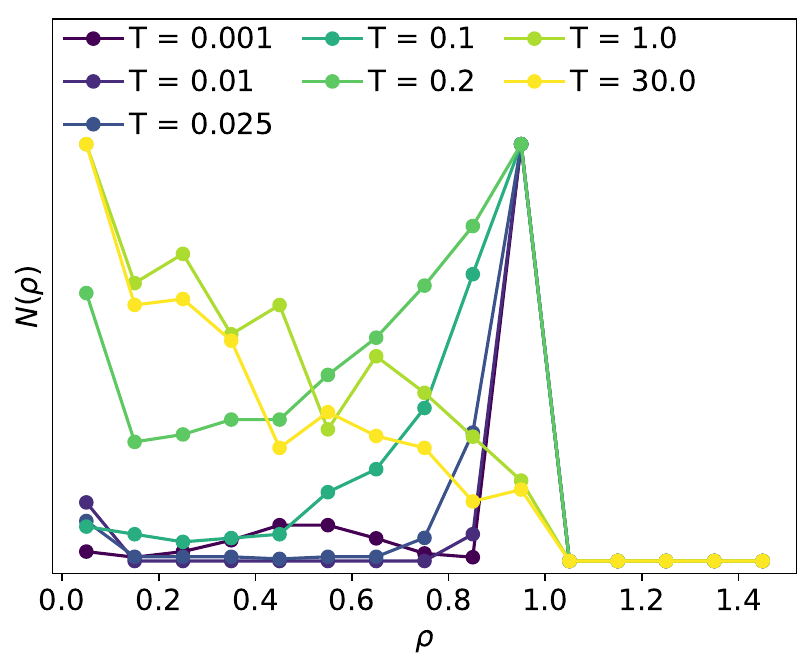}
    \caption[Final histogram of $\rho$ distribution for layered nickelate with $K_\mathrm{oxy}\rightarrow\infty$]{
        \label{nickelate_Koxyinf_rho_histogram.pdf}
        Final histogram, after $10^7$ iterations, of $\rho$ distribution for layered nickelate with $K_\mathrm{oxy}\rightarrow\infty$, at a selection of temperatures. 
        The cell here was a single $10\times30$ nickelate layer.
    }
\end{figure}

\begin{figure}[h]
    \includegraphics[scale=1.0]{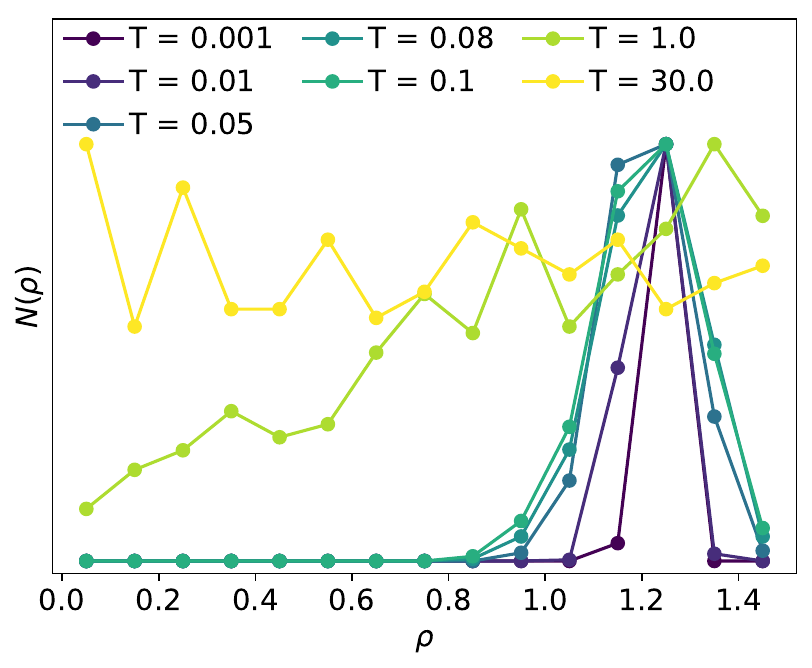}
    \caption[Final histogram of $\rho$ distribution for phase 1 (collinear) perovskite]{
        \label{phase1_rho_histogram.pdf}
        Final histogram, after $10^7$ iterations, of $\rho$ distribution for $8\times8\times8$ phase 1 (collinear) perovskite, at a selection of temperatures. 
    }
\end{figure}

\begin{figure}[h]
    \includegraphics[scale=1.0]{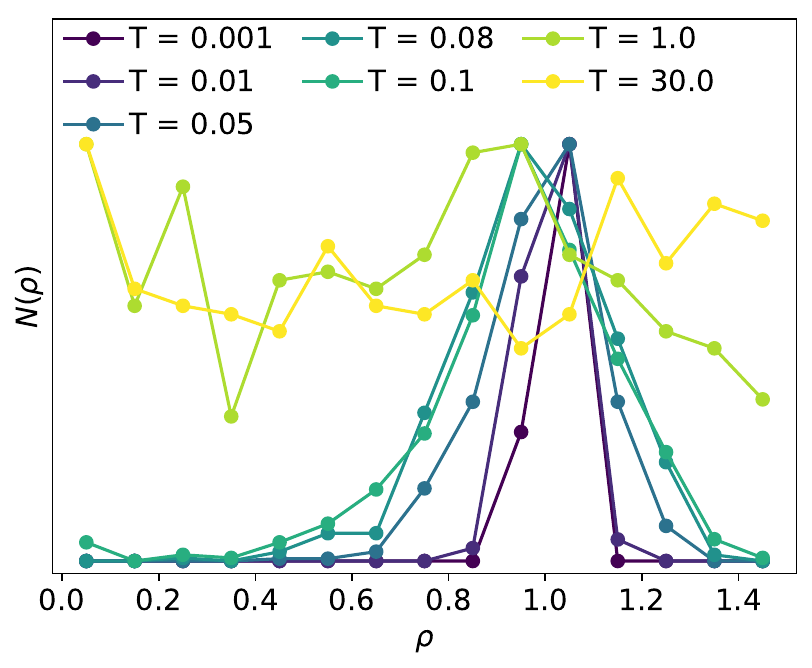}
    \caption[Final histogram of $\rho$ distribution for phase 5 (C-type) perovskite]{
        \label{phase5_rho_histogram.pdf}
        Final histogram, after $10^7$ iterations, of $\rho$ distribution for $8\times8\times8$ phase 5 (C-type) perovskite, at a selection of temperatures. 
    }
\end{figure}

\clearpage
\subsection{\textcolor{black}{Robustness of results with varying ratio of $E_\mathrm{geometry}$ and $E_\mathrm{single-ion}$}}
\label{SI_Section_varying_alpha_magnitude}

\textcolor{black}{In all dynamic-$\rho$ simulations presented in this manuscript, we set $\alpha=-1$ and $\beta=-\alpha/2$. This is a somewhat arbitrary choice intended to keep the energy scales of $E_\mathrm{geometry}$ and $E_\mathrm{single-ion}$ approximately equal.}

\begin{figure}[h]
    \includegraphics[width=\linewidth]{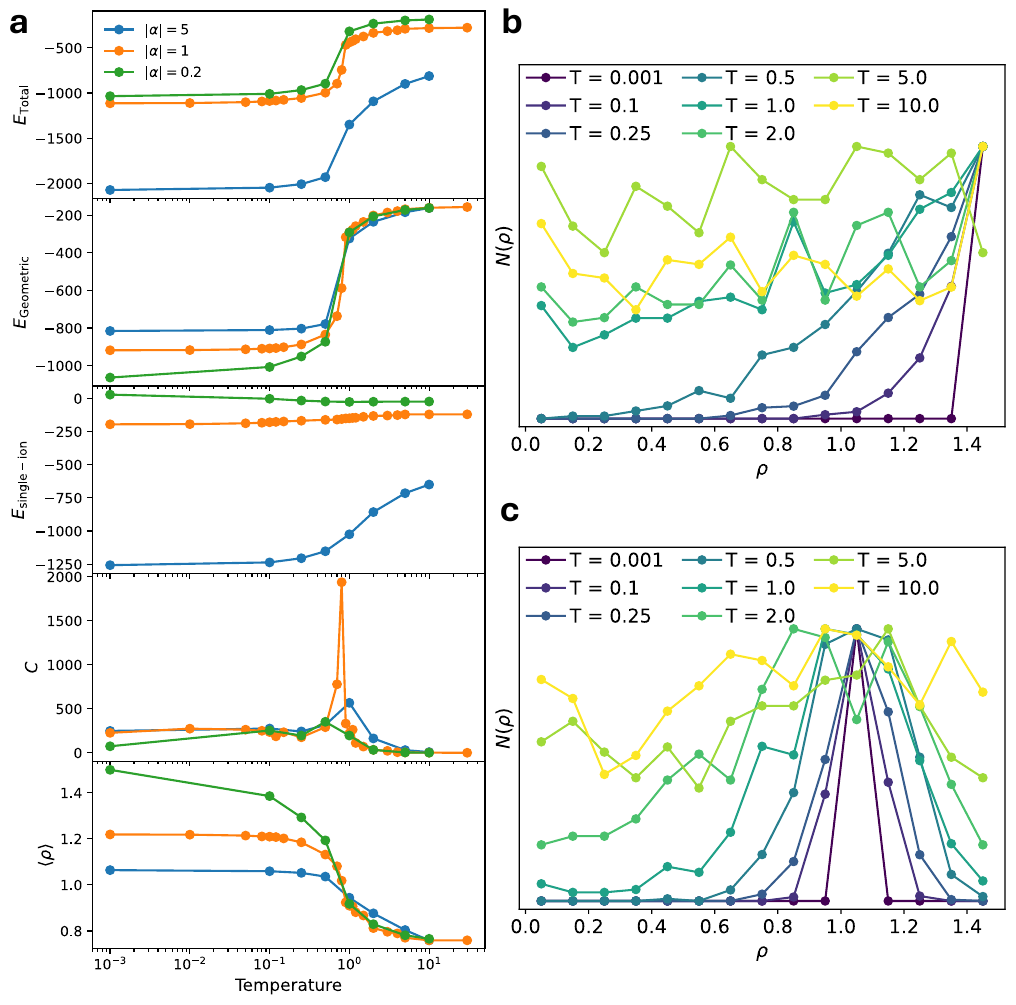}
    \caption[Testing the impact on simulations of much stronger and weaker $E_\mathrm{single-ion}$ relative to $E_\mathrm{geometric}$]{
        \label{phase1_varying-strength-single-ion-test}
        \textcolor{black}{
        Testing the impact on simulations of much stronger $E_\mathrm{single-ion}$ relative to $E_\mathrm{geometric}$. 
        Here, $\alpha=-5,-1/5$ and $\beta=-\alpha/2$, compared a value in the other simulations of $\alpha=-1$. 
        These simulations are run for phase 1 ($J_2=J_1=-1$) perovskite. 
        (a) shows the various observables as a function of simulation temperature. 
        (b,c) Final histogram of $\rho$ distribution at a range of temperatures after $10^7$ iterations, for (b) $\alpha=-1/5$ and (c) $\alpha=-5$. 
        }
    }
\end{figure}

\textcolor{black}{To check the robustness of our findings against this arbitrary choice, we here present alternative simulations for phase 1 ($J_2=J_1=-1$ ). 
In these simulations, we keep the ratio $\beta=-\alpha/2$ constant, but set $|\alpha|$ a factor of 5 larger and smaller [Figure~\ref{phase1_varying-strength-single-ion-test}] than it is set in the normal simulations. 
We keep box size the same as for the simulations in main text (i.e. $8\times8\times8$ supercell of the cubic unit) and run for $10^7$ with simulated annealing as described in the text.}

\textcolor{black}{
We find that results do differ depending on the value of $\alpha$. 
For $\alpha=-0.2$, at low temperatures there is the tendency for $\rho\rightarrow\infty$ due to the domination of the terms favouring Jahn--Teller distortion. This results an accumulation of $\rho$ values at the maximum value allowed by the simulation, which is 1.5. Consequently it is clear that the $E_\mathrm{single-ion}$ term is necessary to ensure stable behaviour and avoid infinite JT distortion. 
For $\alpha=-5$, the opposite occurs, and a JT distortion is rigidly enforced at low-temperatures, as we would expect. 
However, in both cases, application of temperature through the transition results in a distribution of $\rho$ which is consistent with the $\alpha=-1$ case. 
We therefore conclude the value of $\alpha$ we arbitrarily select does not impact our broad conclusions, but is sensible for ensuring reasonable low-temperature behaviour. 
}

\clearpage
\subsection{Non-logarithmic plots of energy with temperature}

\begin{figure}[h]
    \includegraphics[scale=0.7025,trim={0.25cm 0.25cm 0.25cm 0.25cm},clip]{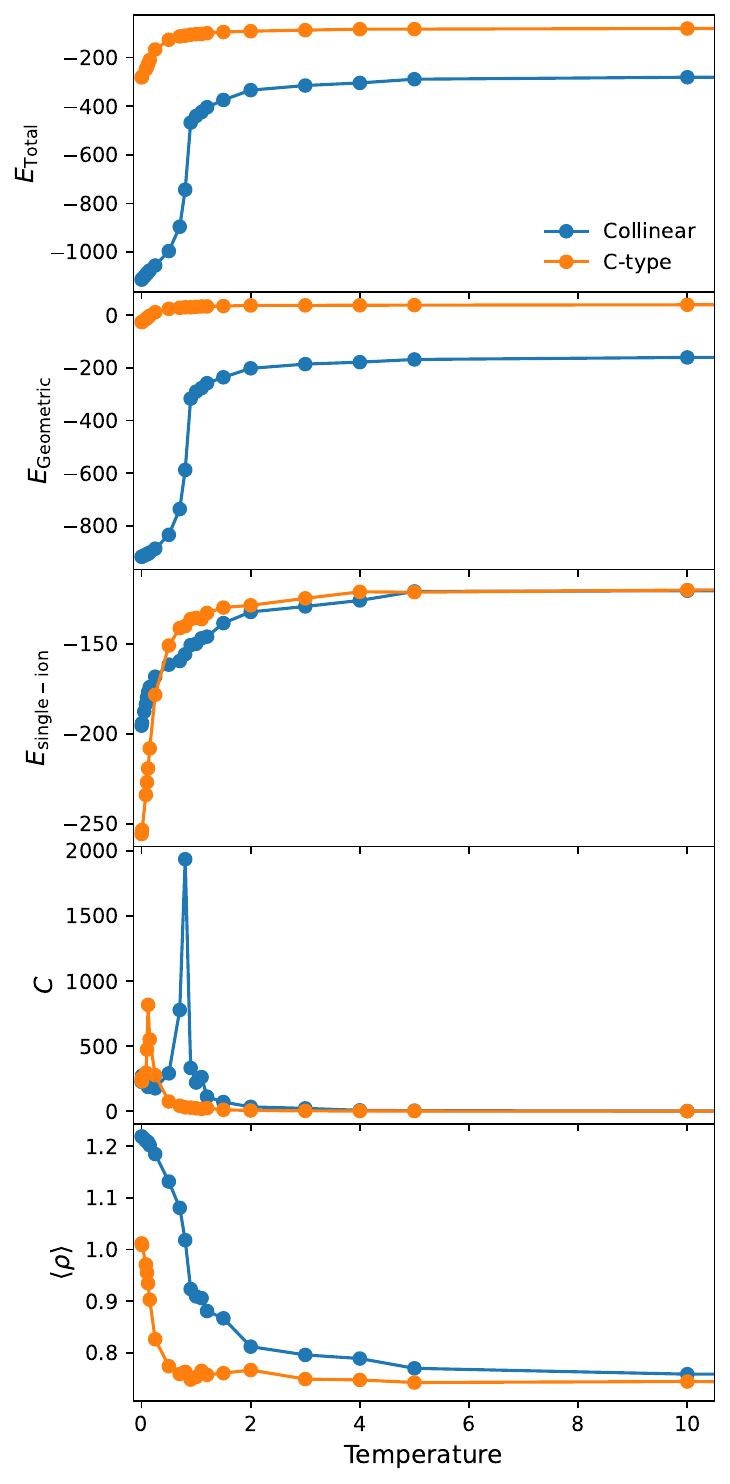}
    \caption[Dependence of heat capacity and energy on temperature for the two perovskite orderings tested, plotted on a non-logarithmic x-axis]{
        \label{perovskite_VT_simulation_nonlog}
        The mean energy terms $E_\mathrm{total}$, $E_\mathrm{geometric}$ (Eq~\ref{Potts_model_modified_perovskite}), and $E_\mathrm{single-ion}$ (Eq~\ref{E_single-ion_term}), heat capacity $C$, and $\langle\rho\rangle$, averaged over the final 10\% of iterations in Monte Carlo simulations, as a function of temperature for the collinear and C-type orbital orderings in an $8\times8\times8$ perovskite lattice. 
        The Monte Carlo simulations ran for $10^7$ iterations in total at each temperature. 
        This figure is reproduced on a logarithmic temperature scale in Figure~\ref{perovskite_VT_simulation.pdf}.
    }
\end{figure}

\begin{figure}[ht]
    \includegraphics[scale=0.7,trim={0.25cm 0.3cm 0.25cm 0.25cm},clip]{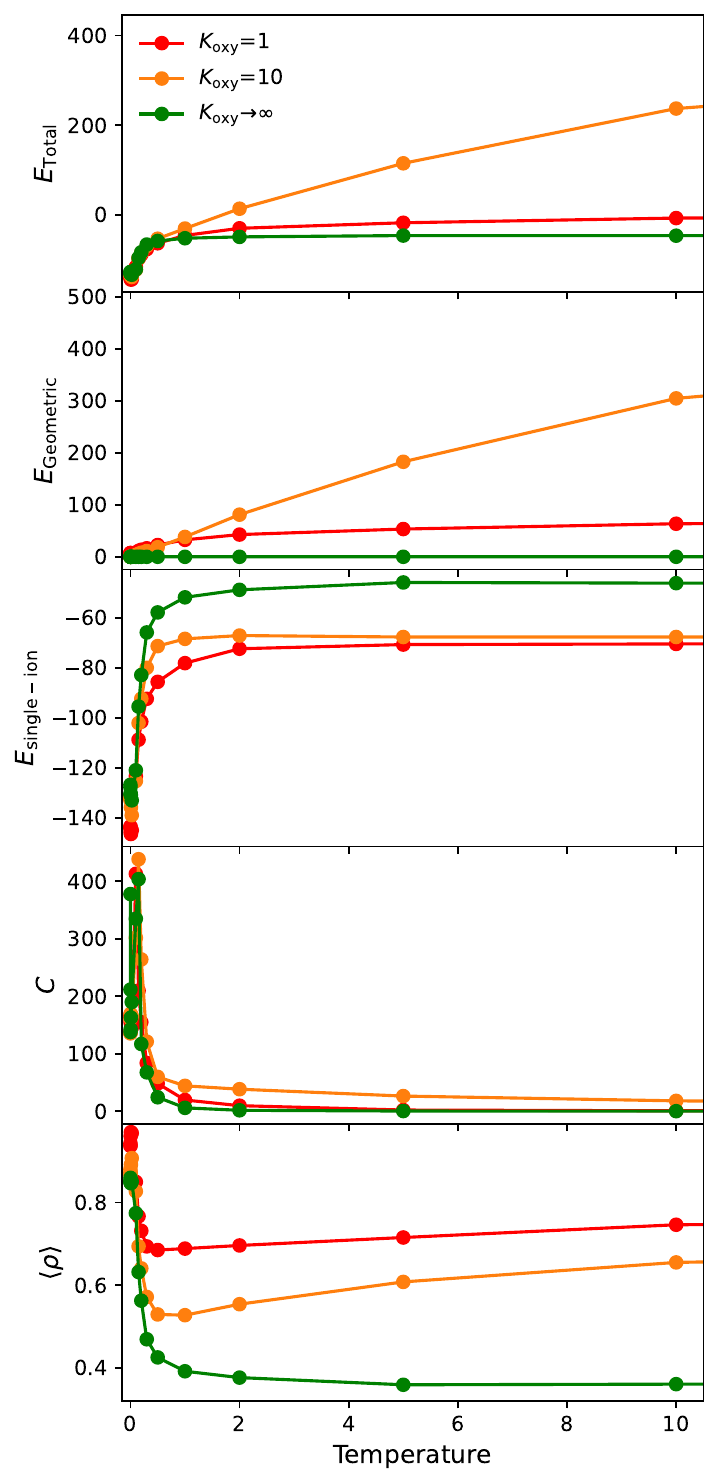}
    \caption[Dependence of heat capacity and energy on temperature 2D layered nickelate lattice, plotted on a non-logarithmic x-axis]{
        \label{nickelate_VT_simulation_not-log}
        The mean energy terms $E_\mathrm{total}$, $E_\mathrm{geometric}$ (Eq~\ref{geometry_term_nickelate}), and $E_\mathrm{single-ion}$ (Eq~\ref{E_single-ion_term}), heat capacity $C$, and mean $\langle\rho\rangle$, averaged over the final 10\% of iterations in Monte Carlo simulations, as a function of temperature for the $10\times30$ nickelate lattice. 
        The Monte Carlo simulations ran for $10^7$ iterations in total at each temperature. 
        We present these results as a function of the strength of the oxygen under-bonding penalty $K_\mathrm{oxy}$. 
        The corresponding figure with a logarithmic temperature scale is presented in Figure~\ref{nickelate_VT_simulation.pdf}.
    }
\end{figure}

\begin{figure}[t]
    \includegraphics[scale=1.22]{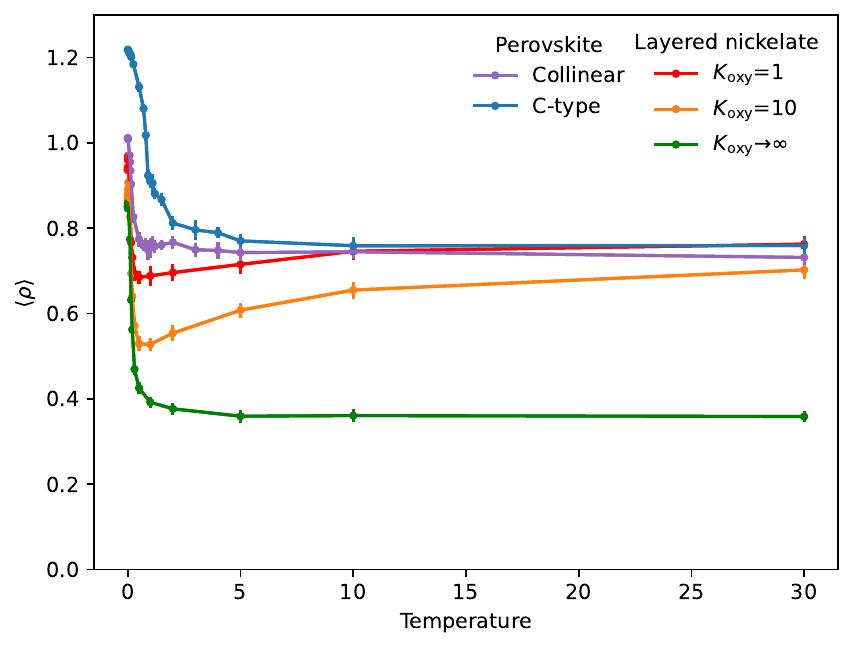}
    \caption[Comparison of $\langle \rho \rangle$ with temperature for the various cases considered on perovskite and 2D nickelate lattices]{
        \label{JT-magnitude-with-temp_non-log}
        $\langle\rho\rangle$ with temperature, comparing phase 1 (collinear) and phase 5 (C-type) perovskite ordering with layered nickelates for various $K_\mathrm{oxy}$. 
        The corresponding figure with a logarithmic temperature scale is Figure~\ref{JT-magnitude-with-temp.pdf}. 
    }
\end{figure}

In the main text, we presented plots of energy, heat capacity, and mean JT magnitude $\langle\rho\rangle$ as a function of temperature, with temperature presented on a logarithmic scale. 
In this section, we present the same figures but on a non-logarithmic scale for reference. 
Figure~\ref{perovskite_VT_simulation_nonlog} shows the results for collinear and C-type orbital ordering perovskites, Figure~\ref{nickelate_VT_simulation_not-log} shows the case for nickelates with $K_\mathrm{oxy}=1,10,10^{10}$, and Figure~\ref{JT-magnitude-with-temp_non-log} shows the combined case.

\clearpage
\subsection{\textcolor{black}{Variable-temperature behaviour at the limits of perovskite phase regions}}
\label{SI_section_check_boundaries_of_phases_VT_behaviour}

\textcolor{black}{
To test the robustness of the variable-temperature behaviour with respect to the choice of $(J_1,J_2)$, we repeated the simulations for phases 1 and 5 at both edges of the corresponding regions of the anisotropic Potts phase diagram, while keeping $\sqrt{J_1^2+J_2^2}$ fixed. 
Specifically, we considered the phase 1/phase 2 ($J_1=\sqrt{2}$, $J_2=0$) and phase 1/phase 6 boundaries ($J_1=1$, $J_2=-1$) for phase 1, and the phase 5/phase 4 ($J_1=1$, $J_2=0$) and phase 5/phase 6 ($J_1\approx0.909$, $J_2\approx-0.427$) boundaries for phase 5. 
}

\textcolor{black}{
In these simulations, we used consistent parameters as in those performed at the centre of the phases. 
For the $E_\mathrm{single-ion}$ term, $\alpha=-1$, $\beta=-\alpha/2=1/2$. 
Simulations ran for $10^7$ iterations. 
}

\begin{figure}[h]
    \includegraphics[width=\linewidth]{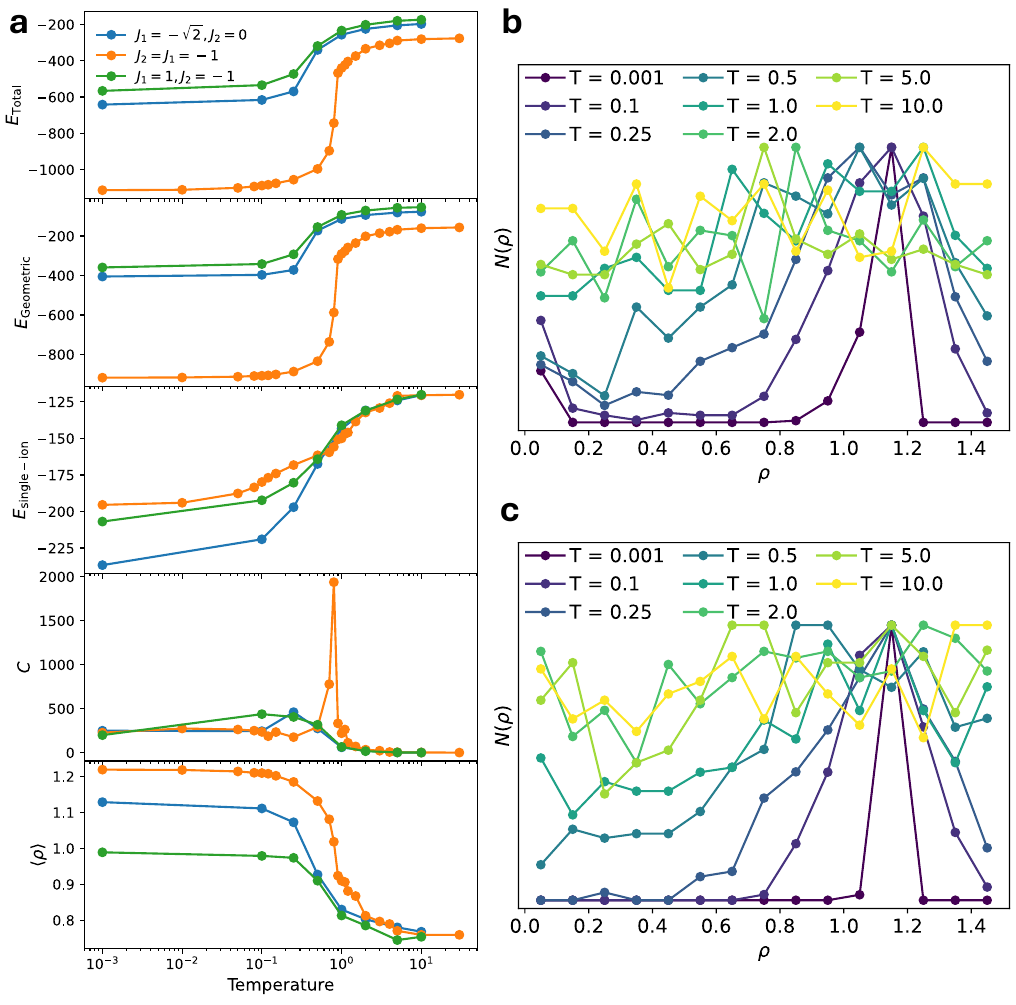}
    \caption[Variable-temperature behaviour at edge points within phase 1 of the perovskite phase diagram]{
        \label{SI-robustness_phase1-boundaries.pdf}
        \textcolor{black}{
        The results of variable-temperature Monte Carlo simulations with dynamic $\rho$ for the perovskite phase 1 region of the anisotropic Potts model, performed close to the phase 1/phase 2 boundary (($J_1=\sqrt{2}$, $J_2=0$) and phase 1/phase 6 boundary ($J_1=1$, $J_2=-1$). 
        The values used in the simulation in the main text ($J_1=J_2=-1$) are also shown for reference. 
        Energy parameters $J_1$ and $J_2$ were chosen to lie near the edge of the phase 1 region while keeping the quadrature sum $\sqrt{J_1^2+J_2^2}$ fixed to the value used for the phase-centre simulations shown in Figure~\ref{perovskite_VT_simulation.pdf}. 
        (a) The mean energy terms $E_\mathrm{total}$, $E_\mathrm{geometric}$, and $E_\mathrm{single-ion}$, heat capacity $C$, and mean $\langle\rho\rangle$ are averaged over the final 10\% of iterations in each simulation. 
        The temperature-dependence is qualitatively consistent with that obtained at the centre of the phase 1 region. 
        (b,c) Histograms of the distribution of $\rho$ for the (b) phase 1/phase 6 boundary and (c) phase 1/phase 2 boundary. 
        }
    }
\end{figure}

\begin{figure}[h]
    \includegraphics[width=\linewidth]{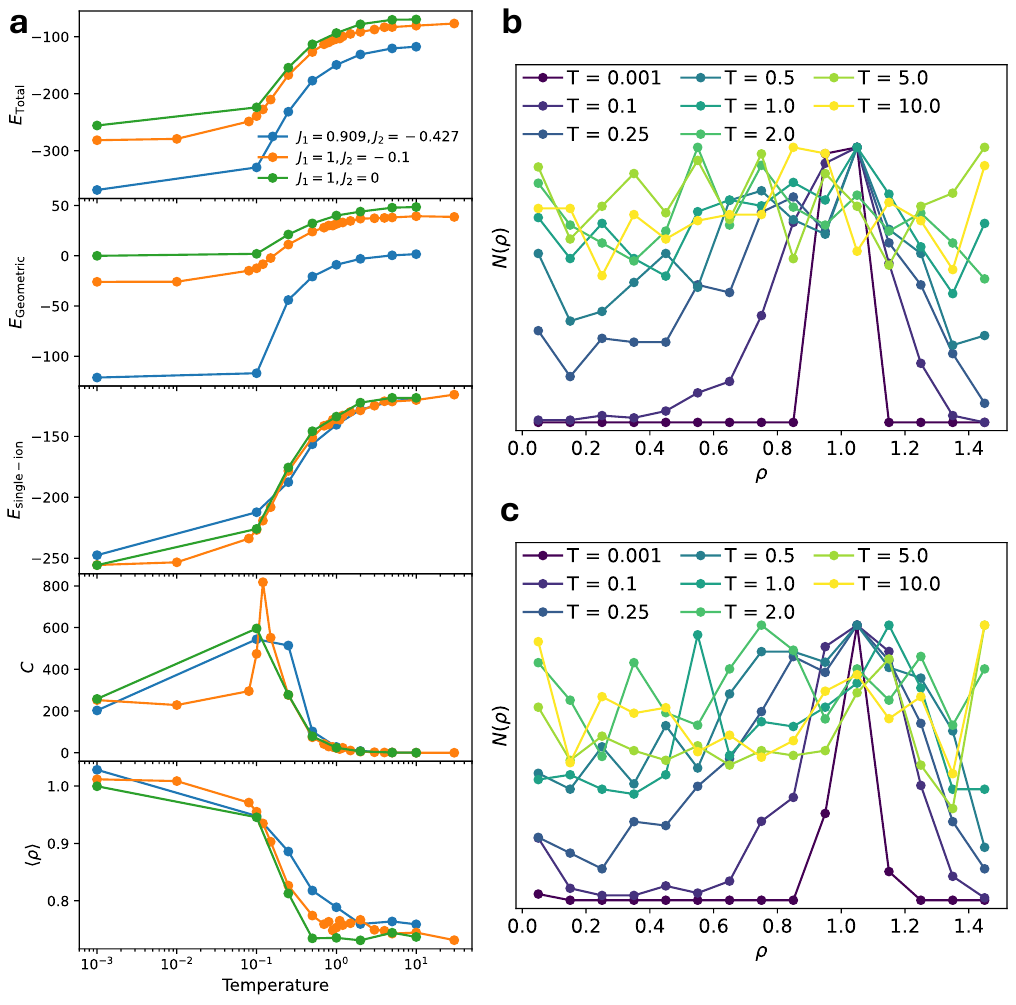}
    \caption[Variable-temperature behaviour at edge points within phase 5 of the perovskite phase diagram]{
        \label{SI-robustness_phase5-boundaries.pdf}
        \textcolor{black}{
        The results of variable-temperature Monte Carlo simulations with dynamic $\rho$ for the perovskite phase 5 region of the anisotropic Potts model, performed close to the phase 5/phase 4 boundary ($J_1=1$, $J_2=0$) and phase 5/phase 6 ($J_1\approx0.909$, $J_2\approx-0.427$) boundary. 
        The values used in the simulation in the main text ($J_1=1$, $J_2=-1/10$) are also shown for reference. 
        Energy parameters $J_1$ and $J_2$ were chosen to lie near the edge of the phase 5 region while keeping the quadrature sum $\sqrt{J_1^2+J_2^2}$ fixed to the value used for the phase-centre simulations shown in Figure~\ref{perovskite_VT_simulation.pdf}. 
        (a) The mean energy terms $E_\mathrm{total}$, $E_\mathrm{geometric}$, and $E_\mathrm{single-ion}$, heat capacity $C$, and mean $\langle\rho\rangle$ are averaged over the final 10\% of iterations in each simulation. 
        The temperature-dependence is qualitatively consistent with that obtained at the centre of the phase 5 region. 
        (b,c) Histograms of the distribution of $\rho$ for the (b) phase 5/phase 4 boundary and (c) phase 5/phase 6 boundary. 
        }
    }
\end{figure}

\textcolor{black}{
As shown in Figures~\ref{SI-robustness_phase1-boundaries.pdf} and \ref{SI-robustness_phase5-boundaries.pdf}, the resulting temperature-dependences are qualitatively consistent with those obtained at the centres of the phase regions. 
}

\textcolor{black}{
We do see that the transition is less pronounced in the $T$-dependence of heat capacity, however, which may be due to the relevant ordering (collinear or C-type) corresponding less decisively to the energy minimum at this position in phase space. This does not qualitatively affect our conclusions
}

\end{document}